\documentclass[12pt]{article}

\usepackage[top=80pt,bottom=85pt,left=60pt,right=60pt]{geometry}
\usepackage{amssymb,amsmath,xypic}
\usepackage{slashed}
\usepackage{graphicx,color,xcolor,subfigure}
\usepackage{float}
\usepackage{sectsty}
\usepackage{cite}
\usepackage{dsfont}
\usepackage{esvect}
\usepackage{amsmath}
\usepackage[debug,pageanchor=false]{hyperref}
\definecolor{link}{rgb}{.8,.15,.1}
\hypersetup{colorlinks=true,linkcolor=link,citecolor=link,urlcolor=link,linktocpage}

 \allowdisplaybreaks

\makeatletter
\@addtoreset{equation}{section}

\usepackage[framemethod=tikz]{mdframed}
\usepackage{tikz-3dplot}

\usepackage{tikz}
\usetikzlibrary{positioning}
\usetikzlibrary{arrows}

\usetikzlibrary{shapes.geometric}
\usetikzlibrary{arrows.meta}
\usetikzlibrary{decorations.pathreplacing,calligraphy}

\tikzset{%
  plus/.pic={
   \node[circle,ball color=orange,inner sep=0pt,minimum width=2.5ex]  {+};
   },
  minus/.pic={
   \node[circle,ball color=green,inner sep=0pt,minimum width=2.5ex]  {-};
   },
    zero/.pic={
   \node[circle,ball color=white,inner sep=0pt,minimum width=2.5ex]  {0};
   }
}

\makeatother

\newcommand{\Vol}{\text{Vol}}
\newcommand{\beq}{\begin{equation}}
\newcommand{\eeq}{\end{equation}}
\newcommand{\bea}{\begin{eqnarray}}
\newcommand{\eea}{\end{eqnarray}}
\newcommand{\nn}{\nonumber}

\def\white1{\textcolor[rgb]{0.98,0.98,0.98}}
\usepackage{rotating}

\begin{document}

\begin{titlepage}

\begin{center}

\vskip .5in 
\noindent

{\Large \bf{Marginally deformed AdS$_5$/CFT$_4$ and 
spindle-like orbifolds}}

		\bigskip\medskip
 Niall T. Macpherson$^{a}$\footnote{macphersonniall@uniovi.es}, Paul Merrikin$^{b}$\footnote{p.r.g.merrikin.2043506@swansea.ac.uk}, Carlos Nunez$^{b}$\footnote{c.nunez@swansea.ac.uk}\\

\bigskip\medskip
{\small 
	
	$a$: Department of Physics, University of Oviedo,\\ 
	Avda. Federico Garcia Lorca s/n, 33007 Oviedo
\\
	and
\\
	 Instituto Universitario de Ciencias y Tecnolog\'ias Espaciales de Asturias (ICTEA),\\ 
	Calle de la Independencia 13, 33004 Oviedo, Spain.\\
~\\
	$b$: Department of Physics, Swansea University, Swansea SA2 8PP, United Kingdom}

	\vskip 3mm

		\vskip 1.5cm 
		\vskip .9cm 
		{\bf Abstract }

		\vskip .1in
	
	\noindent 

{We study marginal deformations of ${\cal N}=2$, $d=4$ long linear quiver CFTs using the holographic description. We find a two-parameter family of AdS$_5$ solutions that generically break all of supersymmetry, but preserve ${\cal N}=1$ for a particular tuning of the parameters. We study the G-structure of the ${\cal N}=2$ ``parent'' and the ${\cal N}=1$ backgrounds and carefully discuss the quantisation of charges in all cases. For  the ${\cal N}=1$ and ${\cal N}=0$ cases, a  picture emerges with ``branes" back-reacted on either a spindle or its higher dimensional analogue. Comments on the marginally deformed dual CFTs are given, together with the study of some observables.}

\end{center}
\vskip .1in

\noindent

\noindent

\vfill
\eject

\end{titlepage}

\tableofcontents
\newpage
\section{Introduction}
 Since twenty-five years ago, the Maldacena conjecture \cite{Maldacena:1997re} motivates the study of CFTs and their associated holographic backgrounds.

Various efforts have focused on classifying families of supergravity backgrounds with an AdS$_{d+1}$ factor. In this work, we are interested on the cases for which families of supergravity solutions have been put in correspondence with families of CFTs. Some examples of these AdS/CFT pairs in dimension $d$ are:  for $d=1$ \cite{Lozano:2020txg,Lozano:2020sae, Lozano:2021rmk}. For $d=2$
\cite{Lozano:2019emq,Lozano:2019zvg,Legramandi:2020txf,Legramandi:2019xqd,Macpherson:2018mif,Lozano:2020bxo}. The case of $d=3$ was studied in \cite{Akhond:2021ffz,Lozano:2016wrs,Assel:2012cj,Assel:2011xz}. The case $d=4$ that occupies us in this work was studied for many different types of dual CFTs. We are specially interested in \cite{Gaiotto:2009gz,Lin:2004nb,Kaste:2003zd,Gauntlett:2004zh,Bah:2019jts, Bah:2022yjf,Macpherson:2016xwk}.
For five dimensional CFTs, families of AdS$_6$ dual geometries were constructed in \cite{Legramandi:2021uds,Uhlemann:2019ypp,Gutperle:2018vdd,Apruzzi:2018cvq,DHoker:2017zwj,DHoker:2017mds}.
The case $d=6$ was studied in \cite{Cremonesi:2015bld,Apruzzi:2015wna,Apruzzi:2014qva, Gaiotto:2014lca,Filippas:2019puw}.
All the above cases preserve supersymmetry. Whilst CFTs in seven dimensions are not compatible with SUSY, non-SUSY  AdS$_8$ backgrounds are found in \cite{Cordova:2018eba}.

As stated, in this paper we are interested in backgrounds with an AdS$_5$ factor. In particular, solutions preserving ${\cal N}=2$ SUSY (eight Poincare SUSYs), ${\cal N}=1$ SUSY (four Poincare SUSYs) and ${\cal N}=0$. The backgrounds of our interest are solutions of type IIA supergravity (or M-theory). Below, we write a new two parameter family of ${\cal N}=0$ backgrounds and a new one-parameter family of ${\cal N}=1$ SUSY backgrounds.
These can be thought of as dual descriptions of marginal deformations of the ``parent'' ${\cal N}=2$ CFTs holographically represented by the Gaiotto-Maldacena backgrounds \cite{Gaiotto:2009gz}.

We focus our attention on these backgrounds in their Type IIA version. These are solutions of cohomogeneity-two, known as `electrostatic backgrounds' (a particular version of the more general cohomogeneity-three M-theory backgrounds).
A picture summarising the families of solutions found, can be seen in Figure \ref{figure1xx}. {Our non-SUSY solutions call for the natural question of stability. In this work, we remain agnostic about this, postponing this to a future study. Nevertheless, two observable quantities we compute suggest stability. In fact, the dynamics of D6 brane probes (sources) --see the analysis below eq.(\ref{eq:needstobezero}) and the masses of spin-two excitations in Section \ref{spintwofluct} display no sign of instability.}

\begin{figure}[h!]
\begin{center}
\begin{tikzpicture}
\draw[-stealth, line width=0.53mm] (-3.3,0)--(3.3,0) node[right ]{$\xi$};
\draw[-stealth, line width=0.53mm] (0,-3.3)--(0,3.3) node[above ]{$\zeta$};
\draw[ blue,line width=0.93mm] (-6,-4.5)--(-5.5,-4.5);
\draw (-5.5,-4.5) node[right]{$ \mathcal{N}=1$ U(1)$_R$ Preserving};
\draw[ red,line width=0.93mm] (2,-4.5)--(2.5,-4.5);
\draw  (2.5,-4.5) node[right]{$ \mathcal{N}=0$ SU(2) Preserving};
\fill[violet!60!magenta] (-0.3,-4.5) circle(.1);
\draw(-0.3,-4.5) node[right]{$~\mathcal{N}=2$};
\clip (-3,-3) rectangle (3,3);
\begin{scope}[cm={0.5,-0.5,  50,50,  (0,0)}]  
\draw[green!70!black,
dashed] (-6,-6) grid (6,6);
\end{scope}

\draw[blue,line width=0.35mm](-3,3)--(3,-3);
\draw[line width=0.5mm,red](-3,0)--(3,0);
\fill[violet!60!magenta] (0,0) circle(.1);
\draw[green!60!black ] (1.45,1.45) node[above] {$\mathcal{N}=0$};
 \node[blue ,rotate=-45] at (1.25,-1.75) {$\zeta=-\xi$};
\end{tikzpicture}
\end{center}
\caption{We see the generic situation (in green dashed lines), for arbitrary values of $(\xi,\zeta)$, the background breaks all SUSY. The solutions along the blue line $\zeta=-\xi$ preserve ${\cal N}=1$ SUSY and consequently a U$(1)_R$-symmetry. The backgrounds parametrised by the line $\zeta=0$ preserve SU$(2)$-isometry (descendent of the original R-symmetry) and SUSY is completely broken. The point $(\xi,\zeta)=(0,0)$ is the infinite family of ${\cal N}=2$ preserving background, with SU$(2)_R\times$ U$(1)_R$ R-symmetry. }
\label{figure1xx}
\end{figure}

Apart from presenting the two-parameter family of solutions, other new results of this work include: a careful discussion of the quantised charges for each family, finding that the {\it spindle} and its higher dimensional version, play a central role for the ${\cal N}=0$ and ${\cal N}=1$ cases. The G-structures associated with ${\cal N}=2$ and ${\cal N}=1$ family of solutions
are given. A field theory analysis of the dual CFTs is also presented. 
A more detailed account of the contents of this paper goes as follows:
\begin{itemize}
\item{In Section \ref{sectionGM} we review the ${\cal N}=2$ Gaiotto-Maldacena system, both in M-theory and in IIA. The new material includes the G-structure and calibration forms (both in ten and eleven dimensions). SUSY preserving probes are studied. The careful calculation of Page charges and the associated balanced quiver field theory are discussed.   }
\item{In Section \ref{eq:deformations} we present the new backgrounds that break SUSY (partially or completely). Whilst in general depending on two parameters labelled $(\xi,\zeta)$, a special situation preserving ${\cal N}=0$ and an SU($2$) isometry arises for $\zeta=0$. Also  backgrounds preserving ${\cal N}=1$ SUSY with an U$(1)_R$ isometry occur for $\zeta=-\xi$. Both cases are carefully analysed. In particular, we find that the standard D6 brane sources of the ${\cal N}=2$ solution are mapped to sources backreacted on a spindle (or its higher dimensional analogue) in the deformed solutions, which leads to a peculiar quantisation condition for the charge of such objects. Stability of some probes is also studied.   }
\item{In Section \ref{secCFT} we discuss some aspects of the CFTs dual to our new backgrounds. In particular, we propose them as marginal deformations of the `parent' ${\cal N}=2$ CFTs. A proposal for the operators deforming the parent theory is given. Consistently with this, the holographic central charge (identified with $a$-central charge) is shown to be the same for all members of the family of solutions. Note that in the limit we work (long linear quivers with large ranks) both CFT central charges  are equal, $a=c$. A mirror-like symmetry relation is proposed between two different quivers. Finally, the equation describing fluctuations of spin-two in the CFT is written, simple universal solutions are presented and a bound on the dimension (mass) of these operators is given.}
\item{Finally, Section \ref{concl} presents conclusions and future directions. Numerous and dense appendices complement the presentation and give many technical details that should be useful for colleagues working on these topics.}
\end{itemize}

\section{Review of Gaiotto-Maldacena and its IIA reduction}\label{sectionGM}
In this section we review a class of ${\cal N}=2$ AdS$_5$ solutions of $d=11$ supergravity found by Gaiotto-Maldacena (GM) \cite{Gaiotto:2009gz} building on the work of \cite{Lin:2004nb}. We will discuss the ${\cal N}=2$ preserving reduction of this class to type IIA supergravity and how it preserves supersymmetry in terms of G-structures. Our main focus will be on a class of solutions first found in \cite{Reid-Edwards:2010vpm}, and further studied and elucidated in \cite{Aharony:2012tz}-\cite{Nunez:2019gbg}.\\
~\\
We begin our discussion with a more general class of ${\cal N}=2$ AdS$_5$ solutions found by Lin-Lunin-Maldacena (LLM). The LLM class has a metric which decomposes as
    \begin{align}
    \frac{ds_{11}^2}{\kappa^{\frac{2}{3}}}&=e^{2\lambda}\bigg[4ds^2(\text{AdS}_5)+y^2 e^{-6\lambda}ds^2(\text{S}^2)+\frac{4}{1-y \partial_y D}(d\tilde{\chi}+A_a dx^a)^2-\frac{\partial_yD}{y}\bigg(dy^2+e^{D}(dx_1^2+dx_2^2)\bigg)\bigg],\nn\\[2mm]
		A_a&=\epsilon_{ab}\partial_{x_b} D,~~~~ e^{-6\lambda}=-\frac{\partial_yD}{y(1- y\partial_yD)},\label{diegoa}
    \end{align}
			where the metrics on AdS$_5$ and S$^2$ have unit radius. The class supports a purely magnetic four-form  $G_4$
		\beq
		G_4= 2\kappa \bigg[(d\chi+ A_a dx^a)\wedge d(y^3 e^{-6\lambda})+ y(1- y^2 e^{-6\lambda}) d A_a\wedge dx^a-\frac{1}{2} \partial_y e^{D}dx^1\wedge dx^2\bigg]\wedge \text{vol}(\text{S}^2),
		\label{diegoar}\eeq
		and its bosonic isometry group is SO(4,2)$\times$\text{SU}(2)$_R\times$\text{U}(1)$_R$ - the latter two factors realising the required R-symmetry of ${\cal N}=2$. The class of backgrounds depends on a single function $D=D(y,x_1,x_2)$ (and its derivatives), satisfying the Toda equation
    \beq
    \nabla_{(x_1,x_2)}^2D+\partial_y^2e^D=0.
    \eeq
~\\
{In what follows, we move from the cohomogeneity-three backgrounds in eqs.(\ref{diegoa})-(\ref{diegoar}), into backgrounds of cohomogeneity-two. This implies that an isometry arises. A special distribution of punctures (D6 branes in Type IIA) generates the  new isometry. The reader should keep in mind that the Gaiotto-Maldacena electrostatic backgrounds are a special case of the more generic ones in eqs.(\ref{diegoa})-(\ref{diegoar}).
}

The GM  class  is defined in terms of the LLM class by introducing new coordinates\footnote{In \cite{Gaiotto:2009gz} the U(1) isometry of the LLM and GM are both confusingly labelled as $\chi$, that these are not actually the same coordinate was pointed out in \cite{Bah:2019jts}. See Appendix \ref{sec:Transformation} for details.}
\beq
x_1=r\cos\beta,~~~~x_2=-r\sin\beta,~~~~\tilde{\chi}=\chi+\beta,\label{eq:newcoords}
\eeq
then imposing that $\partial_{\beta}$ is a U(1) isometry of the metric and fluxes. The  LLM class can then be transformed to the electrostatic form of the GM class (now described by a Laplace equation) via a B$\ddot{\text{a}}$cklund transformation\footnote{One defines new coordinates $(\sigma,\eta)$ to replace $(r,y)$ through $r^2e^D=\sigma^2$, $y=\dot{V}$, $\log r=V'$ where $V=V(\sigma,\eta)$, with the dot and dash defined in \eqref{eq:notation}. See Appendix \ref{sec:Transformation} for details. }. The metric and potential for the 4-form, $G_4=dA_3$, are given by
\begin{align} 
ds_{11}^2&=f_1\Bigg[4ds^2(\text{AdS}_5) +f_2ds^2(\text{S}^2)+f_3d\chi^2+f_4\big(d\sigma^2 +d\eta^2\big)+f_5\Big(d\beta +f_6d\chi\Big)^2\Bigg],\nn\\[2mm]
A_3&=\Big(f_7d\chi +f_8d\beta \Big)\,\wedge\text{vol}(\text{S}^2),\label{eqn:GM}
\end{align}
where the functions $f_i= f_i(\sigma,\eta)$ are all expressed in terms of a single function  $V= V(\sigma,\eta)$ and constant $\kappa$ as
\begin{equation}\label{eq:thefs}
\hspace{-1cm}
\begin{gathered}
        f_1=\kappa^{\frac{2}{3}}\bigg(\frac{\dot{V}\tilde{\Delta}}{2V''}\bigg)^{\frac{1}{3}},~~~~f_2 = \frac{2V''\dot{V}}{\tilde{\Delta}},~~~~~~~~f_3=\frac{4\sigma^2}{\Lambda},~~~~~~~~f_4 = \frac{2V''}{\dot{V}},~~~~~~~~f_5=\frac{2\Lambda V''}{\dot{V}\tilde{\Delta}},
      \\f_6=\frac{2\dot{V}\dot{V}'}{V''\Lambda},~~~~f_7=-\frac{4\kappa \dot{V}^2V''}{\tilde{\Delta}},~~~~~~~f_8=2\kappa\bigg(\frac{\dot{V}\dot{V}'}{\tilde{\Delta}}-\eta\bigg),
        \\ \tilde{\Delta} =\Lambda(V'')^2+(\dot{V}')^2,~~~~~~~~~\Lambda=\frac{2\dot{V}-\ddot{V}}{V''}.
\end{gathered}
\end{equation}
We stress that $\tilde{\Delta}$ should not be confused with $\Delta$ which is defined in later sections. We employ the short hand notation
        \begin{equation}\label{eq:notation}
\dot{V}\equiv \sigma \partial_\sigma V,~~~~~~~~ V' \equiv \partial_\eta V.
    \end{equation}
The solutions that lie within the GM class are defined in terms of the solutions to the following cylindrically symmetric $d=3$ Laplace equation 
\begin{equation}\label{eqn:laplace}
\frac{1}{\sigma}\partial_\sigma (\sigma \partial_\sigma V)+\partial_\eta^2 V\equiv\ddot{V}+\sigma^2V''=0.
\end{equation}
Boundary conditions for this PDE, such that the metric remains regular where the S$^2$ shrinks to zero size (up to $\mathbb{Z}_k$ orbifold singularities), and $\eta$ is bounded to a finite interval $[0,P]$ were found in \cite{Gaiotto:2009gz}. Namely one should have
\beq
\dot{V}\Big|_{\eta=0,P} =0,~~~~\dot{V}|_{\sigma=0} =\mathcal{R}(\eta),\label{eq:boundarycondtions}
\eeq
where ${\cal R}$ is related to the rank function of the dual quiver. This is highly constrained by flux quantisation and by restricting to at most orbifold singularities, namely one should have that
\begin{itemize}
\item ${\cal R}$ is a continuous piece-wise linear function with integer gradient.
\item Any discontinuities in ${\cal R}'$ must happen at integer values of $\eta$.
\item  The gradient of ${\cal R}$ between the discontinuities of ${\cal R}'$ must decrease as one moves towards $\eta=P$. In other words ${\cal R}(\eta)$ is a convex function.
\item $ {\cal R}(0)={\cal R}(P)=0$.
\end{itemize}
When these conditions are satisfied one has $\eta \in [0,P]$ and $\sigma\in [0,\infty)$. For generic values of $\eta$ close to $\sigma=0$, the sub manifold spanned by $(\sigma, \chi)$ vanishes as $\mathbb{R}^2$ in polar coordinates. But if one is at a loci where ${\cal R}'$ is discontinuous instead, the sub-manifold spanned by $(\sigma,\eta,\chi,\beta)$ tends to a $\mathbb{R}^4/\mathbb{Z}_l$ orbifold singularity, where $l$ is the difference between the gradients of ${\cal R}$ on either side of the discontinuity.  To describe the remaining boundaries of the space requires one to define a specific solution, we shall do this from the IIA perspective in Section \ref{sec:nequals2solution}.

\subsection{${\cal N}=2$ preserving reduction to type IIA}\label{eq:neq2class}
Any solution of $d=11$ supergravity with a U(1) isometry $\partial_\psi$ of period $2\pi$ can be reduced to type IIA supergravity through the formulae
\beq
ds^2_{11}= e^{-\frac{2}{3}\Phi}ds^2_{10}+ e^{\frac{4}{3}\Phi}(d\psi+ C_1)^2,~~~A_3= C_3+ B_2\wedge d\psi,\label{eq:reduction}
\eeq
 where $ds^2_{10}$ is the metric in IIA, $\Phi$ the dilaton and the gauge invariant  fluxes are $F_2= dC_1,~ F_4=dC_3-H_3\wedge C_1,~H_3$. The Bianchi identities and equations of motion of type IIA are implied by those of $d=11$ supergravity.

The GM solution in eqs. \eqref{eqn:GM}-\eqref{eq:notation}, has three such U(1) isometries at our disposal, but as shall be made more clear in Section \ref{sec:g-structures}, the full ${\cal N}=2$ supersymmetry can only be preserved when one reduces on $\partial_{\beta}$. In Section \ref{eq:deformations} we  consider other possibilities giving rise to parametric deformations  of this class that break some or all of the ${\cal N}=2$ supersymmetry. The result of reducing the background of eq. \eqref{eqn:GM} along $\partial_{\beta}$ is the following class of solutions in IIA
\beq\label{eq:N=2} 
\begin{aligned}
ds^2&= f_1^{\frac{3}{2}} f_5^{\frac{1}{2}}\bigg[4ds^2(\text{AdS}_5)+f_2ds^2(\text{S}^2)+f_4(d\sigma^2+d\eta^2)+f_3 d\chi^2\bigg],\\[2mm]
e^{\frac{4}{3}\Phi}&=  f_1 f_5,~~~~  H_3 = df_8\wedge \text{vol}(\text{S}^2),~~~~C_1=  f_6d\chi,~~~~ C_3=f_7 d\chi\wedge\text{vol}(\text{S}^2),\nn 
\end{aligned}
\eeq
which, as proven in \cite{Macpherson:2016xwk}, is the most general AdS$_5$ class in IIA  admitting an SU(2) R-symmetry in terms of a round 2-sphere. Note that a positive metric   requires $\frac{\dot{V}}{V''}>0$ except on the boundaries of the space. The Maxwell fluxes are
\beq
F_{2n}=dC_{2n-1}-H_3\wedge C_{2n-3}.
\eeq
While the fluxes $F_n$ are gauge invariant they do not give rise to quantised charges, for that one needs the Page fluxes $\hat F_n$. These are defined in terms of the gauge invariant fluxes through the Poly-form condition 
\beq
\hat F= e^{-B_2}\wedge F,~~~~F=\sum_{n=1}^5 F_{2n},
\eeq 
where $dB_2=H_3$ and the higher fluxes are defined as $F_6=-\star F_4,~F_8=\star F_2,~F_{10}=-\star F_0$. The Bianchi identity of the Maxwell Poly-form $dF-H\wedge F=0$ implies that away from sources the Page fluxes are closed, as such one can define potentials for them - it is not hard to confirm that
\beq
\hat F= d(C\wedge e^{-B_2}),
\eeq
where $C$ is the potential poly-form satisfying\footnote{Note that the general expression is 
\beq
F= dC-H_3\wedge C+ F_0 e^{B_2}\nn,
\eeq
where $F_0$ is the Romans mass, which for us is zero.}
\beq
F= dC-H_3\wedge C\label{eq:potentialpoly}.
\eeq
Defining the NS 2-form potential in terms of an integration constant $k$ we have
\beq
B_2=  \left(2\kappa k+ f_8\right)\text{vol}(\text{S}^2)=2\kappa\left(-(\eta-k)+ \frac{1}{4}\dot{V}f_5f_6\right) \text{vol}(\text{S}^2).\label{eq:neq2B2def}
\eeq
Note that $k$ need not be fixed globally as one traverses the internal space, it can shift due to large gauge transformation - we shall see in the next section that the D6 branes on the boundary $\sigma=0$ demand that $k\in \mathbb{Z}$. In terms of this we find the following Page fluxes
\beq
\hat F_2=d(f_6 )\wedge d\chi,~~~~\hat F_4= 2\kappa d\left(f_6(\eta-k)-2\frac{f_2}{V''f_5}\right)\wedge d\chi\wedge \text{vol}(\text{S}^2),\label{eq:Neq2pagefluxes}
\eeq
which are indeed locally closed.  We  also find it helpful to know the higher Page fluxes, but we  delay presenting them until we can do so concisely in terms of G-structures in the next section.

The existence of $\mathbb{R}^4/\mathbb{Z}_l$ orbifold singularities in the $d=11$ backgrounds, at the points where $\dot{V}$ is discontinuous, and for which the M-theory circle  spanned by $\beta$ vanishes, means that in type IIA we have stacks of $l$ D6 branes at these loci.

\subsection{G-structures, sources and calibrations}\label{sec:g-structures}
In this section we  explain how supersymmetry is preserved in terms of G-structures for LLM, GM and its IIA reduction.\\
~\\
In \cite{Kaste:2003zd}, the G-structure conditions for Mink$_4$ vacua to preserve ${\cal N}=1$ supersymmetry were derived. Namely for a solution of the form
\beq
ds^2= e^{2\hat{A}}ds^2(\text{Mink}_4)+  ds^2(\text{M}_7),
\eeq
where $e^{2\hat{A}}$ and the 4-form $G_4$ have support on  $\text{M}_7$, they take the form
\begin{align}\label{eqn:Geqns}
&d(e^{2\hat{A}} K)=0,~~~~d(e^{4\hat{A}} J)=e^{4\hat{A}} \star_7G_4,\nn\\[2mm]
&d(e^{3\hat{A}} \Omega)=0,~~~~d(e^{2\hat{A} }J\wedge J )=-2 e^{2\hat{A} }G_4\wedge K,
\end{align}
where $K$ is a unit norm real 1-form, $J$ a real 2-form and $\Omega$ a holomorphic 3-form which together span a $d=7$ SU(3)-structure on M$_7$, \textit{i.e.}
\beq
J\wedge \Omega=0,~~~~J\wedge J\wedge J=\frac{3}{4}i\Omega \wedge \overline{\Omega},
\eeq
with $K$ orthogonal to $(J,\Omega)$. These conditions ensure that one can always express $(J,\Omega)$ in terms of a complex vielbein $E^a$ for $a=1,2,3$, that is orthogonal to $K$, as
\beq
J= \frac{i}{2}\left(E^1\wedge \overline{E}^1+E^2\wedge \overline{E}^2+E^3\wedge \overline{E}^3\right),~~~~\Omega= E^1\wedge E^2\wedge E^3,\label{eq:SU(3)decomp}
\eeq
which we make use of below.

In order to express an AdS$_5$ solution in this formalism we make use of the Poincar\'e patch
\beq
ds^2(\text{AdS}_5)=e^{2\rho}ds^2(\text{Mink}_4)+ d\rho^2.
\eeq
G-structure condition for AdS$_5$ do exist \cite{Gauntlett:2004zh}, but we find the Mink$_4$ conditions more convenient for the case at hand as, among other things, they allow us to find supersymmetric embeddings for objects  extended in the Mink$_4$, but not the $\rho$ direction. The G-structure of LLM was computed in this form in \cite{Macpherson:2016xwk}. In terms of the  complex vielbein of \eqref{eq:SU(3)decomp} it is 
\begin{align}\label{eq:GMniceframe}
K&= \kappa^{\frac{1}{3}}e^{-2(\lambda+\rho)}d\left(e^{2\rho}y y_3\right),~~~~E_1= \kappa^{\frac{1}{3}}\sqrt{\frac{-\partial_yD}{y}}e^{\lambda+\frac{1}{2}D}\bigg(dx_1+ i d x_2\bigg),\nn\\[2mm]
E_2&= \kappa^{\frac{1}{3}}e^{-2(\lambda+\rho)}d\left(e^{2\rho}y(y_1+i y_2)\right),\nn\\[2mm]
E_3&= -e^{i \tilde{\chi}}\kappa^{\frac{1}{3}}\frac{2}{\sqrt{1- y \partial_yD}}e^{\lambda}\bigg(d\rho+\frac{1}{2}\partial_y Ddy + i (d\tilde{\chi}+ A_a dx^a)\bigg),
\end{align}
where $y_i$ are a set of embedding coordinates for the unit radius 2-sphere, such that the coordinates on M$_7$ are $(\rho,y,x_1,x_2,\tilde{\chi})$ and the coordinates on S$^2$ -  we correct a typo in  (D.4)
 of \cite{Macpherson:2016xwk} and reinstate the constant $\kappa$. The Mink$_4$ warp factor is
\beq
e^{2A}= 4\kappa^{\frac{2}{3}}e^{2(\rho+\lambda)}.
\eeq
It should be clear that the G-structure forms that follow from the above vielbein are charged under SU(2)$_R\times$U(1)$_R$, \textit{i.e.} the $e^{i \tilde{\chi}}$ factor in $E_3$ is charged under $\partial_{\tilde{\chi}}$, spanning U(1)$_R$, and $y_i$ transform in the \textbf{3} of SU(2)$_R$ which is spanned by the  Killing vectors\footnote{Note that these are dual (on unwarped S$^2$) to the 1-forms  $k_i=\epsilon_{ijk}y_jdy_k$.} on S$^2$. Minimal supersymmetry for Mink$_4$ is 4 real supercharges, this gets doubled due to the SU(2)$_R$ R-symmetry, and doubled again due to the U(1)$_R$ R-symmetry which is how this G-structure realises the 16 real supercharges of ${\cal N}=2$ AdS$_5$ solutions.

After changing the coordinates as in \eqref{eq:newcoords} and performing the B$\ddot{\text{a}}$cklund transformation (see footnote 5 and Appendix \ref{sec:Transformation}) one finds that the resulting G-structure forms for the  GM class can be expressed in terms of the following vielbein on M$_7$
\begin{align}
K&= \frac{\kappa\, e^{-2\rho}}{f_1}d\left(y_3e^{2\rho}\dot{V}\right),~~~~E_1 =  -\sqrt{f_1f_3}\bigg(\frac{1}{\sigma} d\sigma+ d\rho   +i d\chi\bigg),~~~~E_2= \frac{\kappa e^{-2\rho}}{f_1}d\left(e^{2\rho}\dot{V}(y_1+i y_2)\right),\nn\\[2mm]
E_3 &=-e^{i\chi}\sqrt{f_1f_5} \,\Bigg[-\frac{1}{4}f_3 \frac{\dot{V}'}{\sigma} d\sigma -V''d\eta +f_6 d\rho  + i \Big(d\beta +f_6d\chi\Big)\Bigg].\label{eqn:GstructureForms}
\end{align}
The Mink$_4$ warp factor is
\beq
e^{2\hat{A}}= 4e^{2\rho}f_1.
\eeq
Notice that the resulting G-structure forms are still charged under an SU(2)$_R\times$U(1)$_R$ R-symmetry due to the $y_i$ and $e^{i\chi}$ terms respectively, however importantly they are singlets with respect to $\partial_{\beta}$ which allows one to reduce to IIA on this direction without breaking any supersymmetry.

G-structure conditions for Mink$_4$ solutions of type II supergravity were first derived in \cite{Grana:2005sn}: They apply to solutions that decompose in the form
\beq
ds^2= e^{2A}ds^2(\text{Mink}_4)+ ds^2(\text{M}_6),~~~~F= g+ e^{4A}\text{vol}(\text{Mink}_4)\wedge \star_6\lambda(g),
\eeq
where in type IIA supergravity $F=\sum_{n=0}^{5}F_{2n}$ is the RR polyform,  $(A,\Phi,H_3,g)$ have support on M$_6$, and $\lambda(g)=g_0-g_2+g_4-g_6$, where the numerical subscript indicates the degree of the form. G-structure conditions for a subclass of ${\cal N}=1$ Mink$_4$ solution that are sufficient for our purposes are given by
\begin{subequations}\label{eqn:IIAconditions}
\begin{align}
d_{H_3}(e^{3A-\Phi}\Psi_+)&=0,\\
d_{H_3}(e^{2A-\Phi}\text{Re}\Psi_-)&=0,\\
d_{H_3}(e^{4A-\Phi}\text{Im}\Psi_-)&=\frac{e^{4A}}{8}*_6\lambda(g),\label{eqn:Calibrationform}
\end{align} 
\end{subequations}
where the  even/odd form degree bi-linears $\Psi_{\pm}$ are defined in terms of an SU(2)-structure on M$_6$ as
\beq\label{eqn:Psipm}
\Psi_+=\frac{1}{8} e^{\frac{1}{2}z\wedge \overline{z}}\wedge \omega,~~~\Psi_-=\frac{i}{8}  z\wedge e^{-i j},
\eeq
where $(j,\omega)$ are respectively a real and holomorphic 2-form obeying
\beq
j\wedge \omega=0,~~~~j\wedge j=\frac{1}{2}\omega \wedge \overline{\omega},
\eeq
which are orthogonal to the complex vielbein component $z$. Note that supersymmetry imposes $g_6=0$, which means that  $F_{2n}$ are purely magnetic for $n=0,1,2$ and purely electric for $n=3,4,5$, thus the condition \eqref{eqn:Calibrationform} can be used to define a set of canonical potentials for the higher fluxes, \textit{i.e.}
\begin{align}
F_6+F_8+F_{10}&= 8\text{vol}(\text{Mink}_4)\wedge d_{H_3}(e^{4A-\Phi}\text{Im}\Psi_-),\nn\\[2mm]
\Rightarrow C_5+C_7+C_9&=8e^{4A-\Phi}\text{vol}(\text{Mink}_4)\wedge \text{Im}\Psi_-,\label{eq:usefulpotentials}
\end{align}
which we make use of later in the paper.

The analogue of the reduction formula \eqref{eq:reduction} mapping between the G-structures of $d=11$ supergravity and type IIA is
\begin{align}
J &= e^{-\frac{2}{3}\Phi}j  +e^{\frac{1}{3}\Phi}u \wedge (d\psi+C_1),~~~~K=e^{-\frac{1}{3}\Phi} v ,\nn\\[2mm]
\Omega  &=\omega \wedge \Big(e^{-\Phi}u +i(d\psi+C_1)\Big),\label{eqn:11Dto10Dforms}
\end{align}
where we assume that $K$ has no leg in $\psi$, as is the case for eq.\eqref{eqn:GstructureForms} with $\psi=\beta$, and where we decompose  $z=u+i v$.  Decomposing in this way the SU(3)-structure that \eqref{eqn:GstructureForms} leads  to, and  after some massaging, one can express the SU(2)-structure forms implying the supersymmetry of the IIA reduction of GM as
\begin{align}
v&=  \kappa e^{-2\rho}f_5^{\frac{1}{4}}f_1^{-\frac{3}{4}} d\left(e^{2\rho} \dot{V}y_3\right),\nn\\[2mm]
u&= (f_1f_5)^{\frac{3}{4}}\left(\frac{f_3 \dot{V}'}{4 \sigma}+V''d\eta-f_6d\rho\right),\nn\\[2mm]
\omega&=  -2(\kappa)^2 f_1^{-3}f_5^{-\frac{3}{2}}e^{-3\rho}d\bigg(e^{2\rho}\dot{V}(y_1+i y_2)d(e^{i\chi}e^{\rho}\sigma)\bigg),\nn\\[2mm]
j&=\sqrt{f_1 f_5}\left(f_1 f_3 (\frac{d\sigma}{\sigma}+d\rho)\wedge d\chi+ \kappa^2 f_1^{-2}e^{-4\rho}d(e^{2\rho}\dot{V}y_1)\wedge d(e^{2\rho}\dot{V}y_2)\right)\label{eq:neq2iiagstructureforms},
\end{align}
clearly these forms are also charged under SU(2)$_R\times$U(1)$_R$.

Apart from establishing how much supersymmetry a background preserves, G-structures provide useful tools for establishing whether the sources the background has, have a supersymmetric embedding. The sources of interest are D-branes, which have the action
\begin{align}
S_{\text{Dp}}&= S_{\text{DBI}}+S_{\text{WZ}},~~~~S_{\text{DBI}}= T_p\int e^{-\Phi}\sqrt{\det(g+{\cal F})}d^{p+1}\xi,~~~~S_{\text{WZ}}= \mp T_p\int C\wedge e^{-{\cal F}},\label{eq:branactions}
\end{align}
where ${\cal F}=B_2+2\pi \tilde{f}_2$ for $\tilde{f}_2$ a world-volume gauge field, $C$ is the poly-form potential satisfying \eqref{eq:potentialpoly} 
and the pull back onto the Dp brane world volume with coordinates $\xi^{\mu}$ is understood. One takes $-$ for branes and $+$ for anti-branes.  For minimal energy D branes one has that  $S_{\text{DBI}}+S_{\text{WZ}}=0$ when the supergravity equations of motion are satisfied - a supersymmetric embedding implies minimal energy but the converse is not true. A D brane is supersymmetric when it satisfies the $\kappa$-symmetry constraints,  which can be phrased in terms of generalised calibrations \cite{Gutowski:1999tu}, which for Mink$_4$ solutions are themselves defined in terms of the bi-linears $\Psi_{\pm}$ \cite{Martucci:2005ht}. For a Dp brane extended in Mink$_4$ and wrapping some internal cycle  $\Sigma_{p-3}$ this amounts to the requirement that
\beq
e^{4A-\Phi}\sqrt{\det(g+{\cal F})}\bigg\lvert_{\Sigma_{p-3}}d^{p-3}\xi=\pm\Psi^{(\text{cal})}_{\text{Dp}}\bigg\lvert_{\Sigma_{p-3}},~~~\Psi^{(\text{cal})}_{\text{Dp}}\equiv 8e^{4A-\Phi}\text{Im}\Psi_-\wedge e^{-{\cal F}},\label{eq:Dpsusy}
\eeq
where the pull back onto $\Sigma_{p-3}$ is now taken and $\pm$ differentiates between branes and anti-branes - clearly given \eqref{eq:usefulpotentials} this condition implies $S_{\text{DBI}}+S_{\text{WZ}}=0$.

For the IIA reduction of GM we have D6 branes along the boundary $\sigma=0$ at the loci where $\dot{V}$ is discontinuous. Given that these descend from pure geometry in $d=11$ one expects them to be supersymmetric, let us now confirm that. We have the following calibration forms for branes in the GM class  that are extended in $\text{Mink}_4$ with zero world-volume flux ($\tilde{f_2}=0$):
\begin{align}
\Psi^{(\text{cal})}_{\text{D4}}&=e^{4A-\Phi}u,\nn\\[2mm]
\Psi^{(\text{cal})}_{\text{D6}}&=e^{4A-\Phi}(v\wedge j-B_2\wedge u),\nn\\[2mm]
\Psi^{(\text{cal})}_{\text{D8}}&=e^{4A-\Phi}(-\frac{1}{2}u\wedge j\wedge j-B_2\wedge v\wedge j),
\end{align}
with the SU(2)-structure forms defined in \eqref{eq:neq2iiagstructureforms} and $B_2$ in \eqref{eq:neq2B2def} (note that $B_2\wedge B_2=0$). For the D6 branes we should take $\Sigma=(\rho,~\text{S}^2)$ at $\sigma=0$ (note that such branes preserve the symmetry of AdS$_5\times$S$^2$). We find
\begin{align}
\Psi^{(\text{cal})}_{\text{D6}}\bigg\lvert_{(\rho,\text{S}^2)}&=(4\kappa)^3 e^{4\rho}\frac{\dot{V}^2}{V''}d\rho\wedge \text{vol}(\text{S}^2),\\[2mm]
e^{4A-\Phi}\sqrt{\det(g+B_2)}d^{p-3}\xi\bigg\lvert_{(\rho,\text{S}^2)}&=(4\kappa)^3 e^{4\rho}\frac{\dot{V}^2}{V''}\sqrt{1+\frac{\sigma^2 V''}{2 \dot{V}}}\sqrt{1-2 (\eta-k)\frac{\dot{V}'}{\dot{V}}-4 \kappa(\eta-k)^2\frac{V''}{f_7}}d\rho\wedge \text{vol}(\text{S}^2),\nn
\end{align}
which satisfy \eqref{eq:Dpsusy} at the loci $(\sigma=0,\eta=k)$. Given that $k$ appeared first as an integration constant in $B_2$ that can shift due to large gauge transformations (for which $\frac{1}{(2\pi)^2}\int_{\text{S}^2} B_2$ can shift by an integer), we find as expected that unbroken supersymmetry restricts the  D6 branes to lie at integer values of $\eta$ along the $\sigma=0$ boundary, $k$ must be integer for consistency.
%

One might wonder about the supersymmetry of other probe branes extended in AdS$_5$. A quick computation suggests that for D4 branes this is not possible, at least  for the background considered in the next section. In fact, for the solution we consider in Section \ref{sec:nequals2solution} this is certainly the case. For the D8 brane, every term in $\Psi^{(\text{cal})}_{\text{D8}}$  contains one of $(y_3,k_3)$ so is charged under SU(2)$_R$. This means that D8s cannot be added without breaking supersymmetry. We do find however, by taking $y_i=(\cos\phi\sin\theta,\sin\phi\sin\theta,\cos\theta)$ and $\Sigma=(\rho,\phi,\sigma,\eta,\chi)$, that 
\begin{align}
e^{4A-\Phi}\sqrt{\det(g+B_2)}\bigg\lvert_{\Sigma}d^5\xi&=\frac{1}{2}(4\kappa)^4 e^{4\rho}\sigma(2 \dot{V}+\sigma^2 V'')\sin\theta d\rho\wedge d\phi\wedge \sigma \wedge d\eta \wedge d\chi,\\[2mm]
\Psi^{(\text{cal})}_{\text{D8}}\bigg\lvert_{\Sigma}&=\frac{1}{2}(4\kappa)^4 e^{4\rho}\sigma(2 \dot{V}+\sigma^2 V'')\sin^2\theta d\rho\wedge d\phi\wedge \sigma \wedge d\eta \wedge d\chi,\nn
\end{align}
so we can place half BPS D8 branes at $\sin\theta=1$.

\subsection{A particular solution}\label{sec:nequals2solution}
We will be primarily interested in a solution to the Laplace equation (\ref{eqn:laplace}) with the  boundary conditions in eq.\eqref{eq:boundarycondtions} studied  in \cite{Reid-Edwards:2010vpm,Aharony:2012tz,Nunez:2019gbg}. This employs a separation of variables ansatz. In order to be consistent with the conditions presented below eq.\eqref{eq:boundarycondtions} we parametrise the rank functions as
\begin{equation}\label{eqn:Rkoriginal}
\mathcal{R}(\eta) =
    \begin{cases}
     ~~~~~~~~~~~~~~~ N_1\eta & \eta\in[0,1]\\
      N_k+(N_{k+1}-N_k)(\eta-k) & \eta\in[k,k+1]\\
   ~~~~~~~~~   N_{P-1}(P-\eta)  & \eta\in[P-1,P],
    \end{cases}     
\end{equation}
so that the $\eta$ axis is divided into $P$ unit length cells with $k=0,...,P-1$ -  Note that this is not the most general solution possible.   In terms of this rank function one has the following solution
\begin{align}
V(\sigma,\eta) &=-\sum_{n=1}^\infty {\cal R}_n \sin\bigg(\frac{n\pi}{P}\eta\bigg) K_0\bigg(\frac{n\pi}{P}\sigma\bigg),\label{eqn:potential}\\[2mm]
{\cal R}_n&=\frac{2}{P}\int_{0}^P \mathcal{R}(\eta)\sin\bigg(\frac{n\pi}{P}\eta\bigg)d\eta =\frac{2P}{(n\pi)^2}\sum_{k=1}^Pb_k\sin\left(\frac{n\pi k}{P}\right),~~~b_k=2N_k-N_{k+1}-N_{k-1},\nn
\end{align}
where $K_m(\sigma)$ is a modified Bessel function of the second kind and $N_a=0$  for $a=0$ and $a\geq P$. In general one can write the Rank function as a Fourier series
\beq
{\cal R}= \sum_{n=1}^{\infty} {\cal R}_n \sin\left(\frac{n\pi}{P}\eta\right).
\eeq
It is also useful to have an alternative parametrisation of $\dot{V}$, \textit{i.e.} one can show \cite{Reid-Edwards:2010vpm} the equivalence of 
\begin{align}
\dot{V}&=\frac{\pi}{P}\sum_{n=1}^{\infty}{\cal R}_n\sigma\sin\bigg(\frac{n\pi}{P}\eta\bigg) K_1\bigg(\frac{n\pi}{P}\sigma\bigg),\label{eq:alternative}\\[2mm]
&=\frac{1}{2}\sum_{m=-\infty}^{\infty}\sum_{k=1}^Pb_k\left(\sqrt{\sigma^2+(\eta-2m P+k)^2}-\sqrt{\sigma^2+(\eta-2m P-k)^2}\right).\nn
\end{align}
The $m=0$ contribution of the second of these expressions, evaluated at $\sigma=0$, gives rise to the odd extension of $\cal{R}$ defined in the interval $\eta\in [-P,P]$, while the remaining values of $m$ make this $2P$ periodic for $\eta\in \mathbb{R}$. In the $k$'th cell, with $\eta\in[k,k+1]$ we take 
\beq
B^{k}_2=2\kappa\left(-(\eta-k)+ \frac{1}{4}\dot{V}f_5f_6\right) \text{vol}(\text{S}^2),
\eeq
meaning that we perform a large gauge transformation $B_2\to B_2+2\kappa k \text{vol}(\text{S}^2)$ as we traverse between each cell while moving along the $\eta$ axis towards $\eta=P$.\\
~\\
From the perspective of the $d=10$ metric, the coordinates of the Riemann surface are bounded as $\sigma\in [0,\infty)$ and $\eta\in [0,P]$, with the solution regular at all points on the interior\footnote{This follows because \eqref{eqn:laplace} is an elliptic PDE for which extrema of solutions can only lie on the boundaries.}. Let us review the behaviour at the boundaries of the Riemann surface. The reader may also find Appendix \ref{sec:valuesoffi} useful where we summarise the values that the functions $f_i$, appearing in \eqref{eq:N=2}, take at these boundaries.

First we will consider the boundary $\sigma\to \infty$. In the limit $x\to \infty$, $K_0(x)\to\sqrt{\pi} (2x)^{-\frac{1}{2}}e^{-x}$ and so the leading term in \eqref{eqn:potential} is the $n=1$ contribution, this implies that
\beq
V=-{\cal R}_1 e^{-\frac{\pi}{P}\sigma}\sqrt{\frac{P}{2\sigma}}\sin\left( \frac{2\pi}{P} \eta\right) +..,
\eeq
and so the solution reduces at leading order to 
\begin{align}
ds^2&=\kappa \bigg[4\sigma\bigg(ds^2(\text{AdS}_5)+d\chi^2\bigg)+\frac{2P}{\pi}\bigg(d\left(\frac{\pi}{P}\sigma\right)^2+ d\left(\frac{\pi}{P}\eta\right)^2+ \sin^2\left(\frac{\pi}{P}\eta\right)ds^2(\text{S}^2)\bigg)\bigg],\nn\\[2mm]
e^{-\Phi}&=\frac{{\cal R}_1\pi^2}{2 P^{\frac{3}{2}}\sqrt{\kappa}}e^{-\frac{\pi}{P}\sigma}\left(\frac{\pi}{P}\sigma\right)^{-\frac{1}{2}},~~~~H_3=-\frac{4\kappa P}{\pi}\sin^2\left(\frac{\pi}{P}\eta\right) d\left(\frac{\pi}{P}\eta\right)\wedge \text{vol}(\text{S}^2),
\end{align}
where the RR fluxes are zero at leading order and we observe that $(\frac{\pi}{P}\eta,\text{S}^2)$  now span a round, unit radius 3-sphere. This is very similar to the metric found in section  3.9.2 of \cite{DHoker:2017mds}, indeed following a similar argument to that used there\footnote{Specifically one parameterises $ds^2(\text{AdS}_5)= e^{2\rho}\eta_{\mu\nu}dx^{\mu}dx^{\nu}+d\rho^2$ then redefines $(x^{\mu},\rho,\chi)=(4\kappa\sigma)^{-\frac{1}{2}}(\tilde{x}^{\mu},\tilde{\rho},\tilde{\chi})$, which yields $4\kappa\sigma \left(ds^2(\text{AdS}_5)+d\chi^2\right)= \eta_{\mu\nu}d\tilde x^{\mu}d\tilde x^{\nu}+d\tilde \rho^2+d\tilde{\chi}^2$ to leading order in $\sigma$, which is Mink$_6$.} one can show that $4\kappa\sigma \left(ds^2(\text{AdS}_5)+d\chi^2\right)\to ds^2(\text{Mink}_6)$ up to sub-leading terms in $\sigma$. Thus by  defining a new coordinate $ \tilde{r}= e^{-\frac{\pi}{P}\sigma}(\frac{\pi}{P}\sigma)^{-\frac{1}{2}}$ we find to leading order that
\beq
ds^2=ds^2(\text{Mink}_6)+\frac{2P \kappa}{\pi \tilde{r}^2}\bigg(d\tilde{r}^2+\tilde{r}^2ds^2(\text{S}^3)\bigg),~~~H_3=-\frac{4\kappa P}{\pi}\text{vol}(\text{S}^3),~~~e^{-\Phi}=\frac{{\cal R}_1\pi^2}{2 P^{\frac{3}{2}}\sqrt{\kappa}}\tilde{r},
\eeq
which is the near horizon limit of spherically symmetric stack of NS5 branes in flat space. If we tune $2\kappa=\pi$  we find that the charge of these NS5 branes are appropriately quantised, \textit{i.e.}
\beq
Q_{\text{NS5}}=-\frac{1}{(2\pi)^2}\int_{S^3} H_3=P.
\eeq

The limits as $\eta= 0, P$ for $\sigma$ away from its bounds are quite similar, focusing on the former we find that to leading order about $\eta=0$
\beq
\dot V'= f,~~~\dot{V}= f \eta,~~~V''=-\eta\sigma^{-2}\dot f,~~~\ddot{V}=\eta \dot f,~~~f(\sigma)=\frac{\pi^2}{P^2}\sum_{n=1}^{\infty}{\cal R}_n \sigma n^2 K_1\left(\frac{n\pi}{P}\sigma\right),\label{eq:fdef}
\eeq
where $f$ is a positive monotonically decreasing function with finite maximum at $\sigma=0$, which in particular means $|\dot{f}|=-\dot{f}$, the metric in this limit tends to 
\beq
ds^2= \kappa\sqrt{\frac{ 2 f+|\dot f|}{|\dot{f}|}}\bigg[4 \sigma\bigg( ds^2(\text{AdS}_5)+ \frac{|\dot{f}|}{2 f+|\dot{f}|}d\chi^2\bigg)+\frac{2 |\dot{f}|}{\sigma f}\bigg(d\sigma^2+d\eta^2+ \eta^2 ds^2(\text{S}^2)\bigg)\bigg],
\eeq
clearly $(\eta,\text{S}^2)$ is vanishing as the origin of $\mathbb{R}^3$ in polar coordinates at this loci, so the solution is regular - at least away from $\sigma=0$ which is a limit which requires a little more care. One can show $\eta=P$ is likewise regular in a similar fashion.

It is not hard to show that along the entire $\sigma=0$ boundary $\ddot V=0$ to leading order so the solution tends to
\begin{align}
\frac{ds^2}{\kappa}&=\sqrt{\frac{2 \dot{V}}{V''}}\bigg[4ds^2(\text{AdS}_5)+ \frac{2\dot{V}V''}{2\dot{V}V''+(\dot{V}')^2}ds^2(\text{S}^2)+ 2\frac{V''}{\dot{V}}\bigg(d\eta^2+d\sigma^2+ \sigma^2 d\chi^2\bigg)\bigg],\nn\\[2mm]
e^{-4\Phi}&=\frac{V''(2\dot{V}V''+ (\dot{V}')^2)^2}{2^5\kappa^2\dot{V}},~~~~H_3=2\kappa d\left(-\eta+\frac{\dot{V} \dot{V}'}{2 \dot{V} V''+ (\dot{V}')^2}\right)\wedge \text{vol}(\text{S}^2),\nn\\[2mm]
C_1&= \dot{V}'d\chi,~~~~C_3=-4\kappa\left(\frac{\dot{V}^2V''}{2 \dot{V}V''+(\dot{V}')^2}\right) d\chi\wedge \text{vol}(\text{S}^2).
\end{align}
For $0<\eta<P$, and not an integer, we have by definition that at $\sigma=0$, $\dot{V}'$ is constant and $\dot{V}$ is non vanishing, so the sub-manifold spanned by $(\sigma,\chi)$ vanishes as the origin of $\mathbb{R}^2$ as $\sigma\to 0$ provided that $V''$ neither blows up nor vanishes. In the $k$'th cell for $\eta\in(k,k+1)$, using the double series parametrisation in \eqref{eq:alternative}, we find 
\begin{align}
V''&= P_k \label{eqPdef}\\[2mm]
P_k&= \sum_{j=k+1}^P\frac{b_j}{j-\eta}+\frac{1}{2P}\sum_{j=1}^Pb_j\left(\psi\left(\frac{\eta+j}{2P}\right)-\psi\left(\frac{\eta-j}{2P}\right)+\frac{\pi}{2}\left(\cot\left(\frac{\pi(\eta+j)}{2P}\right)-\cot\left(\frac{\pi(\eta-j)}{2P}\right)\right)\right),\nn
\end{align}
for $\psi$ the digamma function. This indeed neither vanishes nor blows up between these bounds so the solution is regular along the $\sigma=0$ boundary when $\eta\notin \mathbb{Z}$. For $\eta=0,P$ the behaviour is analogous, focusing on the former by expanding $(\eta= r\cos\alpha,\sigma=r\sin\alpha)$ for small $r$ we find to leading order
\beq
\dot{V}=N_1 r \cos\alpha,~~~~~~\dot{V}'= N_1,~~~V''=\frac{1}{4P^2}r\cos\alpha\sum_{j=1}^Pb_k\left(2 \psi^1\left(\frac{j}{2P}\right)-\pi^2\csc^2\left(\frac{j\pi}{2P}\right)\right),\label{eqQdef}
\eeq
where $\psi^1$ is the trigamma function, thus the internal metric vanishes as $\mathbb{R}^5$ is polar coordinates at $\sigma=\eta=0$ and the AdS$_5$ warp factor and dilaton are constant, so the solution is again regular at this point - one can show the same is true at $\sigma=\eta-P=0$.  For the final limit $(\sigma=0, \eta= k)$ for $0<k<P$ we expand $(\eta=k- r\cos\alpha,\sigma=r\sin\alpha)$ for small $r$ and find
\beq
\dot{V}=N_k,~~~V''=\frac{b_k}{2r },~~~~\dot{V}'=\frac{b_k}{2}(1+\cos\alpha)+N_{k+1}-N_{k},\label{eq:VsattheD6s}
\eeq
which means the solution at leading order tends to 
\begin{align}
\frac{ds^2}{2\kappa\sqrt{N_k}}&= \frac{1}{\sqrt{\frac{b_k}{r}}}\bigg(4ds^2(\text{AdS}_5)+ ds^2(\text{S}^2)\bigg)+ \frac{\sqrt{\frac{b_k}{r}}}{N_k}\bigg(dr^2+ r^2 ds^2(\tilde{\text{S}}^2)\bigg),~~~e^{-\Phi}=\left(\frac{N_kb_k^3}{2^6\kappa^2r^3 }\right)^{\frac{1}{4}},\nn
\end{align}
which is the near horizon limit of a stack of D6 branes wrapping AdS$_5\times$S$^2$ and where $\tilde{\text{S}}^2$ is spanned by $(\alpha,\chi)$. The flux potentials to leading order are
\beq
B_2=0,~~~C_1=\left(\frac{b_k}{2}(1+\cos\alpha)+N_{k+1}-N_k\right)d\chi,~~~C_3=-2\kappa N_k d\chi\wedge \text{vol}(\text{S}^2),
\eeq
thus the charge of D6 branes is appropriately quantised, \textit{i.e.}
\beq
F_2=-\frac{1}{2}b_k \text{vol}(\tilde{\text{S}}^2)~~~\Rightarrow~~~Q^k_{D6}=-\frac{1}{2\pi}\int_{\tilde{S}^2}F_2= b_k=2N_k-N_{k-1}-N_{k+1}.
\eeq
So the solution has a stack of source NS5 branes at $\sigma=\infty$, D6 branes at $(\sigma=0,\eta=k)$ for $k=1,...P-1$ and is regular everywhere else.

One can also define a Page charge for D4 branes at $\sigma=0$ using \eqref{eq:Neq2pagefluxes}. First off we note that for regular points between $k<\eta<k+1$ we simply have
\beq
\hat F_4= 2\kappa{\cal R}'' (\eta-k)d\eta\wedge d\chi\wedge \text{vol}(\text{S}^2),
\eeq
which is zero at such loci, it is likewise not possible to define a non trivial Page flux at the upper or lower bound of $\eta$. Things fare better at the loci of the D6 branes, we find close to $\eta=k$, where \eqref{eq:VsattheD6s} hold we can integrate on $(\chi,\text{S}^2)$ and the semi circular contour defined by $(\eta=k- r\cos\alpha,\sigma=r\sin\alpha)$ for $r$ infinitesimal and $0\leq \alpha\leq \pi$. The important thing to appreciate is that $\hat F_4$ in \eqref{eq:Neq2pagefluxes} is the Page flux in the $k$'th unit cell, but the contour we are following starts in the $(k-1)$'th  cell and crosses into the $k$'th cell at $\alpha= \frac{\pi}{2}$. Performing the integral carefully one finds
\beq
Q^k_{D4}=-\frac{1}{(2\pi)^3}\int_{\text{S}^2\times \tilde{\text{S}}^2}\hat F_4=N_k-N_{k-1},\label{eq:neq2F4pagecharge}
\eeq
these should be interpreted as colour branes, not flavour ones.
We note that the total charge of D6 and D4 branes obeys
\beq
Q_{D6}=\sum_{k=1}^{P-1}Q^k_{D6}=N_{P-1}+N_1,~~~~
Q_{D4}=\sum_{k=1}^{P-1}Q^k_{D4}=N_{P-1},
\eeq
 The total charge of D4 branes quoted above includes the 'true' colour D4 present in the background, but also the charge of  four-brane induced on the D6 and NS branes. If we are interested only in the 'true' D4 charge, in the interval $[k,k+1]$ there are $N_k$
of them. Also, the total charge of D4-branes is
\begin{equation}
Q_{D4}^{\text{Total}}=\int_0^P R(\eta) d\eta.
\end{equation}
 
 In the next section we will consider supersymmetry breaking deformations of these solutions.

\section{Supersymmetry breaking deformations}\label{eq:deformations}
\subsection{SL(3,$\mathds{R}$) Transformation}
We can dimensionally reduce the GM class of solutions to Type IIA in a more general manner when compared to equations \eqref{eqn:GM} and \eqref{eq:reduction}. Indeed, by first parametrising
\beq
ds^2(\text{S}^2)= d\theta^2+\sin^2\theta d\phi^2,~~~~\text{vol}(\text{S}^2)=\sin\theta d\theta \wedge d\phi,
\eeq
we can perform an SL(3,$\mathds{R}$) transformation amongst the three U(1) directions $(\partial_\beta,~\partial_\chi,~\partial_\phi)$, as follows
   \begin{equation}\label{eqn:SL(3,R)}
 \begin{gathered}
d\beta \rightarrow a\, d\chi+b\, d\beta +c\, d\phi,~~~~d\chi \rightarrow p\, d\chi +\xi\, d\beta + m \,d\phi,~~~~d\phi \rightarrow s\, d\chi +\zeta\,d\beta +u\,d\phi,\\
\begin{vmatrix}
~p&\xi&m~\\
~a&b&c~\\
~s&\zeta&u~\\
\end{vmatrix}
=p(bu-\zeta c)-\xi (au-sc)+m(a\zeta -sb)=1,
\end{gathered}
\end{equation}
with the U(1) component of the SU(2)$_R\times$U(1)$_R$ R-symmetry becoming 
\beq\label{eqn:U(1)}
\begin{aligned}
\text{U(1)}_R &=\chi+\phi \\&\rightarrow (p+s)\chi +(\xi+\zeta)\beta + (m+u)\phi.
\end{aligned}
\eeq
The nine SL(3,$\mathds{R}$) transformation parameters can be  reduced, without loss of generality, to three free parameters (corresponding to the three U(1) directions being mixed). The exact choice of these free parameters provide options when reducing to Type IIA. For our purposes, we simply absorb $(p,b,u)$ into the definitions of $(\chi,\beta,\phi)$, respectively (setting them to one). This avoids re-defining the three U(1)s amongst themselves, and immediately eliminates three of the nine parameters. \\\\
In the case of a dimensional reduction along $\beta$, it proves useful to keep $(\xi,~\zeta)$ free\footnote{Which from the determinant  \eqref{eqn:SL(3,R)}, requires $a=c=0$.}, allowing for the preservation of the $U(1)$ component \eqref{eqn:U(1)} when $\zeta=-\xi$. The determinant in \eqref{eqn:SL(3,R)} now reduces to the condition $ms=0$. Hence, the third free parameter can either be $m$ (with $s=0$), or  $s$ (with $m=0$). For the resulting IIA backgrounds however, one can set both $m=s=0$ without loss of generality\footnote{ With $s$ free (and $m=0$) one gets \eqref{eqn:generalresult1} with $\phi\equiv \phi+s \chi$, and with $m$ free (and $s=0$), one instead has $\chi\equiv \chi +m \phi$. Hence, in both cases, one can set $m=s=0$ without loss of generality, resulting in a two-parameter family of solutions.}. Hence, for the specific 11D coordinate transformations just outlined, we have
\beq\label{eqn:exacttransformation}
d\beta \rightarrow   d\beta  ,~~~~~~~~~~~~~~~~~d\chi \rightarrow d\chi +\xi\, d\beta  ,~~~~~~~~~~~~~~~~~~d\phi \rightarrow d\phi + \zeta\,d\beta.
\eeq
{Notice that the parameters $(\xi,\zeta)$ should take integer values to avoid spoiling the periodicity of the angles $(\beta,\chi,\phi)$.}
The result of reducing to type IIA on $\partial_{\beta}$, using \eqref{eq:reduction}, is the family of backgrounds
\begin{align}
        ds^2&=e^{\frac{2}{3}\Phi}f_1\bigg[4ds^2(\text{AdS}_5)+f_2d\theta^2+f_4(d\sigma^2+d\eta^2)\bigg]+f_1^2e^{-\frac{2}{3}\Phi} ds^2_2,\nn\\[2mm]
    ds^2_2 &= f_3f_5 d\chi^2 + \sin^2\theta f_2\bigg[f_3 (\xi d\phi-\zeta d\chi )^2 + f_5\Big( -\zeta f_6 d\chi+(\xi f_6+1)d\phi\Big)^2\bigg]\nn
        \\[2mm]
e^{\frac{4}{3}\Phi}&= f_1 f_5\bigg[  (1 +\xi f_6)^2 +\xi^2 \frac{f_3}{f_5}+ \zeta^2 \frac{f_2}{f_5} \sin^2\theta\bigg] ,~~~~ B_2  = \sin\theta \,\bigg[\zeta  f_7 d\chi  - (f_8+\xi f_7 ) d\phi \bigg] \wedge d\theta,\nn\\[2mm]
       C_1&= f_1 f_5 e^{-\frac{4}{3}\Phi} \bigg[\bigg( f_6(1 + \xi f_6) +\,\xi\frac{ f_3}{f_5} \bigg)d\chi +  \zeta \sin^2\theta \frac{f_2}{f_5}   d\phi \bigg],~~~~ C_3=f_7  d\chi \wedge \text{vol}(\text{S}^2)\label{eqn:generalresult1}.
\end{align}
From eq. \eqref{eqn:exacttransformation}, it is clear to see that $\zeta\neq 0$ breaks the S$^2$ of the resulting Type IIA background (and hence the SU(2)$_R$ component of the R-Symmetry). Hence, when $\zeta=-\xi\neq0$, because the U(1)$_R$ component \eqref{eqn:U(1)} is independent of the reduction coordinate $\beta$, the resulting background preserves $\mathcal{N}=1$ Supersymmetry (with a U(1)$_R$ R-symmetry). When both parameters are zero, the $\mathcal{N}=2$ case \eqref{eq:N=2} is recovered (preserving the full R-Symmetry). In all other cases, the Supersymmetry is broken completely. A summary of this discussion is given in Table \ref{table:1} and in Figure \ref{figure1xx}.
   
    \begin{table}[h!]
     \begin{center}
\begin{tabular}{c | c c c c  }
$\beta$- Reduction&$\mathcal{N}$&U(1)$_R$&SU(2)$_R$  \\
\hline
$\xi=\zeta=0$&$ 2$ &$\checkmark$&$\checkmark$ \\
$\xi=-\zeta\neq 0$&$ 1$ &$\checkmark$&$\times$  \\
$\xi\neq 0,~\zeta=0$&$ 0$ &$\times$&$\checkmark$  \\
$\xi=0,~\zeta\neq0$&$ 0$ &$\times$ &$\times$  \\
$\xi\neq0,~\zeta\neq 0$&$ 0$ &$\times$&$\times$ 
\end{tabular}
\end{center}
\caption{In this table we see the different possible reductions in $\beta$ in terms of the two relevant parameters $(\xi,\zeta)$. The quantity ${\cal N}$ indicates the amount of SUSY preserved. We also indicate which part of the R-symmetry $\text{SU}(2)_R\times \text{U}(1)_R$ is inherited in the background. }
\label{table:1}
\end{table}
We will first focus on the SU(2)$\times$U(1)  preserving $\mathcal{N}=0$ reduction given in Table \ref{table:1}, obtained by fixing $\zeta=0$ in equation \eqref{eqn:generalresult1}. After that we investigate the $\mathcal{N}=1$ case (with $\zeta=-\xi$).

\subsection{SU(2)$\times$U(1) preserving $\mathcal{N}=0$ deformation }
In this section we will study the unique deformation of the solution of Section \ref{sec:nequals2solution} which preserves none of the supersymmetry  while retaining  SU(2)$\times$U(1) isometry, see Table \ref{table:1}. Whilst the SU$(2)$ isometry descends from the SU$(2)_R$ part of the R-symmetry of the ${\cal N}=2$ backgrounds, the U$(1)$ does not originate in the U$(1)_R$ of the parent backgrounds.

Fixing $\zeta=0$ in eq.\eqref{eqn:generalresult1}, we find the solution can be succinctly written as
 \begin{align}
ds^2&= f_1^{\frac{3}{2}} f_5^{\frac{1}{2}}\sqrt{\Delta}\bigg[4ds^2(\text{AdS}_5)+f_2ds^2(\text{S}^2)+f_4(d\sigma^2+d\eta^2)+\frac{f_3}{\Delta} d\chi^2\bigg],~~~~e^{\frac{4}{3}\Phi}=  f_1 f_5\Delta, \label{N=0metric}\\[2mm]
\Delta&=(1+\xi f_6)^2+ \frac{\xi^2 f_3}{f_5},~~~~  H_3 = d(f_8+\xi f_7)\wedge \text{vol}(\text{S}^2),\nn\\[2mm]
C_1&=  \frac{f_6+\xi\left(\frac{f_3}{f_5}+f_6^2\right)}{\Delta}d\chi,~~~~ C_3=f_7 d\chi\wedge\text{vol}(\text{S}^2),\nn 
\end{align}
which is clearly a parametric deformation of eq.\eqref{eq:N=2}, reducing to it exactly when $\xi=0$. In the $k$'th cell, with $\eta\in[k,k+1]$ we now take
\beq
B^{k}_2=2\kappa\left(-(\eta-k)+ \frac{1}{4}\dot{V}f_5f_6-\xi\dot{V} f_2\right) \text{vol}(\text{S}^2),
\eeq 
which leads to the 4-form Page flux
\beq
\hat F_4= 2\kappa d\left(\frac{f_6(\eta-k)-\frac{2 f_2}{V'' f_5}+\xi\left(\dot{V}f_2 f_6-\left(\frac{f_3}{f_5}+f_6^2\right)\left(-(\eta-k)+\frac{1}{4} \dot{V}f_5 f_6\right)\right)}{\Delta}\right)\wedge d\chi\wedge\text{vol}(\text{S}^2).\label{eq:neq0hatF4def}
\eeq
The various functions $f_i(\eta,\sigma)$ are defined in eq.\eqref{eq:thefs}, and we will focus on the particular solution with $V$ defined as in eq.\eqref{eqn:potential} or equivalently eq.\eqref{eq:alternative}.\\
~~\\
To better understand this deformation it is instructive to study how $\xi\neq 0$ modifies the  behaviour of the ${\cal N}=2$ solution at the boundaries of the space.
{We remind the reader that the parameters $(\xi,\zeta)$ should take values in the integers to avoid spoiling the periodicity conditions of the angles in eq.(\ref{eqn:exacttransformation}).}

Indeed, at  $\sigma=\infty$  $f_6$, $f_7=-2\kappa\dot{V}f_2$ and  $f_3 f_5^{-1}$ all tend to zero, which makes $\xi$ drop out of the solution. This means that the deformed solutions tend to the undeformed one as we approach this boundary, so again there are $P$ NS5 branes at $\sigma=\infty$. Similarly at $\eta= 0,P$, but for $\sigma\neq 0$, $f_3$ and $f_7$ tend to zero while the remaining $f_i$ are nowhere zero or infinite so the behaviour, while modified, is qualitatively the same as the ${\cal N}=2$  solution, namely the solution is regular with the sub-manifold spanned by $(\eta,\text{S}^2)$ vanishing as  $\mathbb{R}^3$ in polar coordinates. The more note worthy modification happens along the $\sigma=0$ boundary:

At a generic point along $\sigma=0$, with $\eta\in (k,k+1)$,  $f_3$ tends to zero while $f_5$ is finite and $f_6=\dot{V'}=N_{k+1}-N_{k}$. As such we find
\beq
\Delta\to  \left(1+\xi (N_{k+1}-N_{k})\right)^2.
\eeq
Thus the sub-manifold spanned by $(\sigma,\chi)$, ignoring a constant multiplicative factor, tends to 
\beq
d\sigma^2+\frac{\sigma^2}{l_k^2}d\chi^2, ~~~~l_k=1+\xi (N_{k+1}-N_{k}),\label{eq:orbifold}
\eeq
while the rest of the space is finite, non zero, and independent of $\sigma$ at this loci. As such the regular zero one gets at generic points along the $\sigma=0$ boundary, when $\xi=0$, becomes a $\mathbb{R}^2/\mathbb{Z}_{l_k}$ orbifold singularity when $\xi\neq 0$, as long as $\xi$ is an integer.

To approach $(\sigma=0,\eta=0)$ we define $(\eta=r\cos\alpha,\sigma=r\sin\alpha)$ and expand about $r=0$. Since $f_3$ vanishes with $f_5,f_6$ constant we find 
\beq
\Delta\to (1+\xi N_1)^2.
\eeq
As such we again find that the $(\sigma,\chi)$ directions reproduce orbifold behaviour like \eqref{eq:orbifold}, with $l_0= 1+\xi N_1$. However the $(\sigma,\chi)$ coordinates are part of a larger space at this loci, when $\xi=0$ they combine with the rest of the internal space to give the origin of $\mathbb{R}^5$ in polar coordinates. When $\xi\neq0$, rather than a regular zero, we find a  $\mathbb{R}^5/Z_{l_0}$ orbifold singularity. The behaviour about $(\sigma=0,\eta=P)$ is analogous giving rise to a $\mathbb{R}^5/Z_{l_{P-1}}$ orbifold.

The most interesting modification to the behaviour happens at $(\sigma=0,\eta=k$), for $k$ the loci of a stack of D6 branes when $\xi=0$. Expanding again $(\eta=k- r\cos\alpha,\sigma=r\sin\alpha)$ for  $r\sim 0$ we find the metric and dilaton tend to 
\begin{align}
\frac{ds^2}{2\kappa\sqrt{N_k}}&= \sqrt{\Delta_k}\bigg[\frac{1}{\sqrt{\frac{b_k}{r}}}\bigg(4ds^2(\text{AdS}_5)+ ds^2(\text{S}^2)\bigg)+ \frac{\sqrt{\frac{b_k}{r}}}{N_k}\bigg(dr^2+ r^2 \left(d\alpha^2+\frac{\sin^2\alpha}{\Delta_k} d\chi^2\right)\bigg)\bigg],\nn\\[2mm]
e^{-\Phi}&=\left(\frac{N_kb_k^3}{2^6\kappa^2r^3 }\right)^{\frac{1}{4}}\Delta_k^{-\frac{3}{4}},~~~B_2=-2\kappa N_k\text{vol}(\text{S}^2),\label{eq:neq0D6s}\\[2mm]
C_1&=  \frac{\xi k^2b_k^2\sin^2\alpha+ g(\alpha)(1+\xi g(\alpha))}{\Delta_k}d\chi,~~~C_3=-2\kappa N_k d\chi\wedge \text{vol}(\text{S}^2)\nn,
\end{align}
where we define the functions
\beq
\Delta_k= \frac{1}{4}\xi^2b_k^2  \sin^2\alpha+\left(1+\xi g(\alpha)\right)^2,~~~g(\alpha)=\cos^2\left(\frac{\alpha}{2}\right)(N_k-N_{k-1})+\sin^2\left(\frac{\alpha}{2}\right)(N_{k+1}-N_{k})\label{eq:usefulfunction}.
\eeq
At the poles of the deformed 2-sphere spanned by $(\alpha,\chi)$, we have
\beq
\Delta_k(\alpha=0)=l^2_{k-1},~~~~\Delta_k(\alpha=\pi)=l^2_{k},\label{eq:usefulfunctionatpoles}
\eeq
with $\Delta_k$ finite and non zero between these bounds. If we had $l_k=l_{k-1}=1$ the deformed 2-sphere would become a round one, however as $\xi b_k= l_{k-1}-l_{k}$, for $b_k$ the charge of the D6 brane stack at $\eta=k$ when $\xi= 0$, we necessarily have $l_k\neq  l_{k-1}$ leading to $\mathbb{R}^2/\mathbb{Z}_{l_{k-1}}$ and $\mathbb{R}^2/\mathbb{Z}_{l_{k}}$ conical singularities at the respective poles
for $\xi\neq 0$. This is the behaviour of a so called ``spindle'', which is the weighted projective space $\mathbb{WCP}^1_{[l_{k-1},l_{k}]}$. Spindles have the  topology of a  2-sphere with orbifold singularities at the poles, specifically $\mathbb{WCP}^1_{[n_-,n_+]}$ has $\mathbb{R}^2/\mathbb{Z}_{n_{\mp}}$ orbifold singularities at the south/north poles with $n_->n_+$ and gcd$(n_-,n_+)=1$. Such orbifolds have received a lot of attention recently in the context of the near horizon limit of D branes which wrap them and their dual CFTs . See the papers 
\cite{Ferrero:2020twa,Ferrero:2020laf,Hosseini:2021fge,Boido:2021szx,Ferrero:2021wvk,Bah:2021hei,Ferrero:2021ovq,Couzens:2021rlk,Faedo:2021nub,Ferrero:2021etw,Couzens:2021tnv,Couzens:2022yjl,Arav:2022lzo,Couzens:2022yiv,Suh:2022pkg,Amariti:2023gcx,Inglese:2023tyc,Faedo:2024upq} for examples of works in which the spindle manifold plays a central role in holographic duals. This is not the situation we find here, instead we find the behaviour of D6 branes extended in (AdS$_5$,~S$^2$) and back-reacted on a cone whose base is $\mathbb{WCP}^1_{[l_{k-1},l_{k}]}$. The charge of the D6 branes is given by 
\beq
Q^k_6=-\frac{1}{2\pi}\int_{\mathbb{WCP}^1_{[l_{k-1},l_{k}]}}F_2=-\frac{1}{2\pi}\int_{\chi=0}^{\chi=2\pi} C_1\bigg\lvert_{\alpha=0}^{\alpha=\pi}= \frac{2N_k-N_{k+1}-N_{k-1}}{l_{k}l_{k-1}},\label{quantcond1}
\eeq
yielding precisely the rational quantisation condition one should get when integrating (over the spindle) the field strength of the connection of a U(1) orbifold bundle over a spindle \cite{Ferrero:2021etw}. This follows because the Euler characteristic on the spindle is itself rational, {\it i.e.} we find
\beq
\chi_E=\frac{1}{2\pi}\int_{\mathbb{WCP}^1_{[l_{k-1},l_{k}]}}R \text{vol}_2= \frac{l_{k-1}+l_k}{l_{k-1}l_k}=2-\left(1-\frac{1}{l_k}\right)-\left(1-\frac{1}{l_{k-1}}\right),
\eeq
where $\text{vol}_2$ is the volume form on $\mathbb{WCP}^1_{[l_{k-1},l_{k}]}$.

Given the $d=11$ origin of this solution, that we find the behaviour of D6 branes back-reacted on a spindle should not be surprising. Indeed starting from the following embedding of the Taub-Nut metric into $d=11$ 
\beq
ds^2=ds^2(\mathbb{R}^{1,6})+  h\left(dr^2+ r^2(d\alpha^2+\sin^2\alpha d\chi^2)\right)+ \frac{1}{h}(d\beta+N \cos\alpha d\chi)^2,~~~~h=1+\frac{M}{r}\nn,\label{eq:tndef}
\eeq
we can produce the metric and 2-form very similar to those in eq.\eqref{eq:neq0D6s} (but with D6  branes extended on $\mathbb{R}^{1,6}$ )  by first performing the coordinate transformation $\beta\to \beta+\frac{1}{2}(N_{k+1}-N_{k-1})\chi$ followed by $\chi\to\chi+\xi \beta$,  then reducing to IIA on $\partial_{\beta}$. This reproduces the singular behaviour of \eqref{eq:neq0D6s} close to $r=0$, but with AdS$_5\to \text{Mink}_5$, one should identify $M=b_k$. This solution likewise preserves no supersymmetry for $\xi\neq0$.

As in the $\xi=0$ limit, a Page charge of D4 branes at the loci of the D6 branes can be defined using eq. \eqref{eq:neq0hatF4def}. Integrating carefully as described above eq. \eqref{eq:neq2F4pagecharge} we find
\beq
Q_4^k=-\frac{1}{(2\pi)^2}\int_{\text{S}^2\times \mathbb{WCP}^1_{[l_{k-},l_k]}}F_4= \frac{N_k}{l_k}-\frac{N_{k-1}}{l_{k-1}},
\eeq
which are again colour charges. 
We note that the total charge of D6 and D4 branes obey
\beq
Q_{D6}=\sum_{k=1}^{P-1}Q^k_{D6}=\frac{N_{P-1}+N_1}{l_0 l_P},~~~~
Q_{D4}=\sum_{k=1}^{P-1}Q^k_{D4}=\frac{N_{P-1}}{l_P}.
\eeq

While it is no great surprise that this deformation contains D6 branes, as supersymmetry is now broken it is no longer guaranteed that they are stable. A stable D brane configuration should have minimal energy, the action of such a brane must satisfy 
\beq
S= S_{\text{DBI}}+S_{\text{WZ}}=0,\label{eq:needstobezero}
\eeq
on shell, where the DBI and WZ actions are defined in \eqref{eq:branactions}. To establish whether this is true for the D6 branes in the solution at hand we need to construct the higher form potentials $C_7,C_5$, whose pull back onto $\text{AdS}_5\times \text{S}^2$ appears in the WZ action. To proceed we note that when $\xi=0$ supersymmetry is recovered so we know the form that $C_7,C_5$ should take in this limit from \eqref{eq:usefulpotentials}, the $\xi\neq 0$ limit must be a parametric deformation of this. We find 
\begin{align}
\left(C_7-B^k_2\wedge C_5\right)\bigg\lvert_{\xi=0}&=\frac{1}{2}(4\kappa)^3\bigg[\frac{2\dot{V}^2}{V''}\text{vol}(\text{AdS}_5)\wedge \text{vol}(\text{S}^2)- e^{4\rho}\dot{V}\text{vol}(\text{Mink}_4)\wedge \text{vol}(\text{S}^2)\wedge (\sigma d\sigma)\nn\\[2mm]
&+2 \sigma^4\text{vol}(\text{AdS}_5)\wedge d\chi\wedge d(\sigma^{-2}\dot{V}\cos\theta)-2 \sigma e^{4\rho}\text{vol}(\text{Mink}_4)\wedge d\chi\wedge d\sigma\wedge d(\dot{V}\cos\theta)\nn\\[2mm]
&-(\eta-k)\bigg(\frac{4\sigma^2 f_6}{f_3}\text{vol}(\text{AdS}_5)-e^{4\rho}\text{vol}(\text{Mink}_4)\wedge \left(\dot{V}'d\sigma-\sigma^2\partial_{\sigma}(\sigma^2\dot{V})d\eta\right)\bigg)\wedge \text{vol}(\text{S}^2) \bigg],\nn
\end{align}
where it is actually only the very first term that is relevant for the D6 branes. Since $\hat F=d(C\wedge e^{-B_2})$ and it is not hard to establish that  $\hat F_8$ contains only order 0, 1 and 2 terms in $\xi$, clearly we must have 
\beq
C_7-B^k_2\wedge C_5=\left(C_7-B^k_2\wedge C_5\right)\bigg\lvert_{\xi=0}+\frac{1}{2}(2\kappa)^3\left(\xi X_7+\xi^2 Y_7+ Z_7\right),
\eeq
with $Z_7$  containing terms of any order in $\xi$, but necessarily closed. Consistency of this ansatz with $\hat F_8$ fixes
\begin{align}
Y_7&=  e^{4\rho}\dot{V}^2 \text{vol}(\text{Mink}_4)\wedge \text{vol}(\text{S}^2)\wedge d\dot{V},\\[2mm]
X_7&=\frac{1}{2}e^{4\rho} \text{vol}(\text{Mink}_4)\wedge \text{vol}(\text{S}^2)\wedge\Bigg[d\left(\frac{\dot{V}^2 \dot{V}'}{V''}\right)-\frac{2\dot{V}^2}{f_2}d\eta\\[2mm]
&~~~~~~~~~~~~~~~~~~~~~~~~~~~~~~~~~~~~~~~~~~~~+(\eta-k)\left(d\left(\frac{\dot{V}\dot{V}'}{V''}-\ddot{V} \dot{V}\right)+ 4 \dot{V}(\dot{V}'d\eta- \sigma V''d\sigma)\right)\Bigg]\nn,
\end{align}
and we shall decompose $Z_7$ in terms of an arbitrary function $p=p(\eta,\sigma)$ as
\beq
Z_7=\left( p \,\text{vol}(\text{AdS}_5)\wedge \text{vol}(\text{S}^2)+\frac{1}{4}e^{4\rho} \text{vol}(\text{Mink}_4)\wedge \text{vol}(\text{S}^2)\wedge dp\right),
\eeq
such that it is manifestly closed and contains only forms on the external space whose exterior derivatives respect the isometries of Mink$_4$. The WZ action of a D6 brane of world volume (AdS$_5$,S$^2$)  then takes the form
\beq
S_{\text{WZ}}= (4\kappa)^3T_6\int \left(\frac{\dot{V}^2}{V''}+\frac{p}{2} \right)\text{vol}(\text{AdS}_5)\wedge \text{vol}(\text{S}^2).
\eeq
We find to leading order about $(\sigma=0,\eta=k)$ that
\beq
e^{-\Phi}\sqrt{\det(g+ B_2)}\bigg\lvert_{(\text{AdS}_5,~ \text{S}^2)}=(4\kappa)^3\sin\theta\frac{\dot{V}^2}{V''}\xi\sqrt{2\dot{V}V''\Delta_k}.
\eeq
We can thus fix $p$ such that  \eqref{eq:needstobezero} is satisfied for a D6 brane at $(\sigma=0,\eta=k)$,
\beq
p=2 \xi\frac{\dot{V}^2}{V''}\sqrt{\frac{\Delta}{f_5}},
\eeq
achieves the desired goal, but so does the sum of this and any function tending to zero at the loci of the D6 branes.

\subsection{U(1)$\times$U(1) preserving ${\cal N}=1$ deformation}\label{sec:neq1sol}
In this section we will study the unique deformation of the solution of section \ref{sec:nequals2solution} which preserves ${\cal N}=1$ supersymmetry while retaining a U(1)$\times$U(1) isometry.\\
~\\
Fixing  $\zeta=-\xi\neq0$ in \eqref{eqn:generalresult1}, we find  that we can write the solution in the following form
\begin{align}
ds^2&= f_1^{\frac{3}{2}} f_5^{\frac{1}{2}}\sqrt{\Xi}\bigg[4ds^2(\text{AdS}_5)+f_4(d\sigma^2+d\eta^2)+ds^2(\text{M}_3)\bigg],~~~~e^{\frac{4}{3}\Phi}=  f_1 f_5\Xi\label{eq:neq1sol}\\[2mm]
H_3 &= df_8\wedge \text{vol}(\text{S}^2)+ \xi \sin\theta df_7\wedge d\theta\wedge (d\phi+d\chi),\nn\\[2mm]
C_1&=  \frac{\left(f_6+\xi\left(f_6^2+\frac{f_3}{f_5}\right)d\chi-\xi \frac{f_2}{f_5}\sin^2\theta d\phi\right)}{\Xi},~~~~ C_3=f_7 d\chi\wedge\text{vol}(\text{S}^2),\nn 
\end{align}
where we introduce the functions
\beq
\Xi=\Delta+\xi^2\frac{f_2}{f_5}\sin^2\theta,~~~\Delta=(1+ \xi f_6)^2+\xi^2\frac{f_3}{f_5},~~~\Pi=1+\xi^2f_2\frac{f_3+f_5 f_6^2}{f_3 f_5}\sin^2\theta.
\eeq
The 3-manifold M$_3$ can be expressed in two ways which are useful
\begin{align}
ds^2(\text{M}_3)&=f_2\left(d\theta^2+\frac{\Delta}{\Xi}\sin^2\theta D\phi^2\right)+\frac{f_3}{\Delta} d\chi^2=f_2\left(d\theta^2+\frac{1}{\Pi}\sin^2\theta d\phi^2\right)+\frac{\Pi}{\Xi}f_3D\chi^2,\nn\\[2mm]
D\phi&= d\phi+\frac{\Delta-1-\xi f_6}{\Delta}d\chi,~~~D\chi=d\chi+\frac{\Pi-1+\frac{f_2 f_6}{f_3}\sin^2\theta}{\Pi}d\phi.\label{eqn:M3}
\end{align}
Clearly $\Xi=\Delta=\Pi=1$ when $\xi=0$, so we have another parametric deformation of the ${\cal N}=2$ solution. We define the NS 2-form in the $k$'th cell to be 
\beq
B_2=2\kappa\left(-(\eta-k)+ \frac{1}{4}\dot{V} f_5 f_6-\xi \dot{V} f_2\right)\sin\theta d\theta\wedge d\phi- \xi\dot{V} f_2\sin\theta d\theta\wedge d\chi,
\eeq
 from which it follows that the Page flux of D4 branes in the $k$'th cell is given by
\beq
\hat F_4= 2\kappa d\left(\frac{(\eta-k)\left(f_6+\xi\left(\frac{f_3}{f_5}+f_6^2\right)\right)-2 \frac{f_2}{V'' f_5}(1+\xi f_6)}{\Xi}\right)\sin\theta d\theta\wedge d\phi\wedge d\chi.
\eeq

 We will now discuss the global properties of the solution with $\xi\neq 0$.\\
~\\
First we note that for generic values of $(\eta,\sigma)$, $\Xi,\Delta,\Pi$ are finite and non zero for all $\theta$ with $\Pi\to 1$ at the poles of the deformed S$^2$ spanned by $(\theta,\phi)$. Thus the second expression for M$_3$ makes it clear that the deformed 2-sphere still behaves as a round S$^2$ topologically\footnote{Note that we can parametrise $\sin\theta=x$ for $x$ small at the poles, one then finds $D\chi\to d\chi+ q(\eta,\sigma)x^2d\phi$, where $q$ is easily determined. One can then send $D\chi\to d\chi$ up to leading terms in $x$ through $\chi\to \chi-\frac{1}{3}q x^3$, so the fibration is topologically trivial at the poles.}.

It is not hard to establish that as $\sigma\to\infty$ the $\xi$ dependence drops out of the solution, making it tend to the ${\cal N}=2$ solution. As such there are again $P$ NS5 branes at the $\sigma=\infty$ boundary. Likewise at $\eta=0,P$ (but $\sigma \neq 0$), because $f_2,f_7,f_8$ tend to zero while the remaining $f_i$ are finite and non zero, we find that $\Pi\to 1$, $D\chi\to d\chi$, $f_2\to \eta^2 f_4$, and that $\Xi$ is a finite non zero function of $\sigma$ - thus from the second parametrisation of M$_3$ we see that the $(\eta,\theta,\phi)$ direction vanish as $\mathbb{R}^3$ in polar coordinates, yielding a regular zero again. As with the previous deformation the more interesting modified behaviour with respect to $\xi=0$ happens  along the $\sigma=0$ boundary:

At generic points along the $\sigma=0$ boundary, away from the loci of the D6 branes when $\xi=0$, the solution again contains orbifold singularities. For $\eta\in (k,k+1)$ we find that
\beq
\Delta\to  l_k^2,~~~~\Xi\to l_k^2+ \frac{1}{2}\xi^2 {\cal R}V''\sin^2\theta,~~~D\phi\to d\phi+\frac{\xi(N_{k+1}-N_k)}{l_k}d\chi ~~~~l_k=1+\xi(N_{k+1}-N_{k}),
\eeq  
making $\Xi$ a finite nowhere vanishing function of $(\eta,\theta)$ and the connection of $D\phi$ topologically trivial. Using the first expression for M$_6$ we then see clearly  that the $(\sigma,\chi)$ directions yield a $\mathbb{R}^2/\mathbb{Z}_{l_k}$ orbifold singularity. 

We once more study the $\eta=\sigma=0$ limit by defining $(\eta=r\cos\alpha,\sigma=r\sin\alpha)$ for small $r$. Since $f_3$ vanishes with $f_5,f_6$ finite we have
\beq
\Xi\to \Delta\to l_{0}^2,
\eeq
it is then not hard to see from the first parametrisation of M$_3$ that the internal space is vanishing as $\mathbb{R}^5/\mathbb{Z}_{l_0}$ with the external space finite, just as was the case at this loci for the ${\cal N}=0$ deformation. The behaviour at $\eta-P=\sigma=0$ is qualitatively the same, only with a $\mathbb{R}^5/\mathbb{Z}_{l_{P-1}}$ conical singularity. 

The behaviour at $(\sigma=0,~\eta=k)$ for $k$ the loci of a D6 brane when $\xi=0$ is a little subtle. This is because generically the dominant term in $\Xi$ is that containing $f_2 f_5^{-1}$, the exception is when we are also at one of the  poles of the $(\theta,\phi)$ deformed S$^2$, where more care is required. Assuming first that we are not at a pole we expand $(\eta=k-r \cos\alpha,~\sigma=r \sin\alpha)$ in small $r$ and find to leading order that
\beq
\Xi\to\frac{b_k \xi^2 N_k\sin^2\theta}{4r},~~~\Delta\to \Delta_{k},
\eeq
where $\Delta_k$ is the smooth nowhere zero function defined in \eqref{eq:usefulfunction} tending to  $l_{k-1}^2,~l_{k}^2$ at the respective poles of the deformed $(\alpha,\chi)$ 2-sphere. The solution tends to
\begin{align}
ds^2&=\xi \kappa \sin\theta\bigg[N_k\bigg(4 ds^2(\text{AdS}_5)+ d\theta^2\bigg)+ 4b_k\bigg(dz^2+ z^2 ds^2(\mathbb{B}_3)\bigg)\bigg],~~~~e^{4\Phi}= \kappa^2 \xi^6 N_k^2 \sin^6\theta,\nn\\[2mm]
ds^2(\mathbb{B}_3)&=\frac{1}{4}\bigg(d\alpha^2+\frac{\sin^2\alpha}{\Delta_k}d\chi^2\bigg)+ \frac{\Delta_k}{\xi^2 b_k^2}\left(d\phi+{\cal A}_k\right)^2,~~~~{\cal A}_k=\frac{\Delta_k-1-\xi g(\alpha)}{\Delta_k}d\chi,\nn\\[2mm]
H_3&=-2\kappa \sin\theta d\eta\wedge d\theta\wedge d\phi ,~~~C_1=-\frac{1}{\xi}d\phi,~~~C_3=-2\kappa N_k \sin\theta d\theta\wedge d\phi\wedge d\chi,
\end{align}
where $g(\alpha)$ is defined in \eqref{eq:usefulfunction} and $z=r^2$. Ignoring the overall $\sin\theta$ term, the sub-manifold spanned by $(z,\mathbb{B}_3)$ is a cone of base $\mathbb{B}_3$, while the rest of the space has constant warping. The sub-manifold $\mathbb{B}_3$ is clearly a U(1) fibration over $\mathbb{WCP}^1_{[l_{k-1},l_k]}$, {\it i.e.} $\mathbb{B}_3$  has the following behaviour
\begin{align}
ds^2(\mathbb{B}_3)\bigg\lvert_{\alpha\sim 0}&=\frac{1}{4}\bigg(d\alpha^2+\frac{\alpha^2}{l_{k-1}^2}d\chi^2\bigg)+ \frac{l_{k-1}^2}{\xi^2 b_k^2}\left(d\phi+\frac{\xi(N_k-N_{k-1})}{l_{k-1}}d\chi\right)^2,\nn\\[2mm]
ds^2(\mathbb{B}_3)\bigg\lvert_{\alpha\sim \pi}&=\frac{1}{4}\bigg(d\alpha^2+\frac{(\pi-\alpha)^2}{l_{k}^2}d\chi^2\bigg)+ \frac{l_{k}^2}{\xi^2 b_k^2}\left(d\phi+\frac{\xi(N_{k+1}-N_{k})}{l_{k}}d\chi\right)^2,\nn\\[2mm]
-\frac{1}{2\pi}\int_{\mathbb{WCP}_{[l_{k-1},l_k]}}d{\cal A}&= \frac{\xi b_k}{l_{k-1}l_k},
\end{align}
consistent with this claim. Similar 3 dimensional orbifolds were recently considered in \cite{Inglese:2023tyc} in the context of supersymmetric localisation. Note that the deformed Taub-Nut space discussed around \eqref{eq:tndef}, precisely reproduces the cone over $\mathbb{B}_3$ we find in this limit, it is thus the orbifold singularity associated to this generalised space.

Finally we consider the behaviour as we approach $(\sigma=0,~\eta=k,~\sin\theta=0)$, which can be disentangled by defining
\beq
\eta=k-\rho \cos\alpha\sin^2\mu,~~~~\sigma=\rho \sin\alpha \sin^2\mu,~~~~\sin\theta=2\sqrt{\frac{b_k \rho}{N_k}}\cos\mu,
\eeq
and  expanding in small $\rho$. We find that  this gives rise to a nowhere zero smooth function $\tilde{\Xi}_k$ through
\beq
\sin^2\mu\,\Xi\to \tilde{\Xi}_k=\Delta_k\sin^2\mu+  b_k^2 \xi^2 \cos^2\mu,
\eeq
such that the entire solution tends to
\begin{align}
\frac{ds^2}{2\kappa\sqrt{N_k}}&= \sqrt{\tilde{\Xi}_k }\bigg[\frac{4}{\sqrt{\frac{b_k}{\rho}}}ds^2(\text{AdS}_5)+\frac{\sqrt{\frac{b_k}{\rho}}}{N_k}\bigg(d\rho^2+4 \rho^2 ds^2(\mathbb{B}_4)\bigg)\bigg],~~~e^{-\Phi}=\left(\frac{b_k^3 N_k}{\kappa^2\tilde{\Xi}_k^3\rho^3}\right)^\frac{1}{4},\nn\\[2mm]
B_2&=- b_k \kappa \xi \cos^2\mu d\rho\wedge(d\phi+d\chi),~~~~C_3=- 4 b_k \cos^2\mu d\rho\wedge d\phi\wedge d\chi,\nn\\[2mm]
C_1&=\frac{-b_k^2\xi \cos^2\mu d\phi+\frac{1}{\xi}(\Delta_k-1-\xi g(\alpha)+\frac{b_k^2 \xi}{4} \sin^2\alpha)\sin^2\mu d\chi}{\tilde{\Xi}_k},\label{eq:interestingmonopole}
\end{align}
where the 4 manifold is defined as
\beq
ds^2(\mathbb{B}_4)=d\mu^2+ \frac{1}{4}\sin^2\mu\left(d\alpha^2+\frac{\sin^2\alpha}{\Delta_k}d\chi^2\right)+ \frac{\sin^2\mu\cos^2\mu \Delta_k}{\tilde{\Xi}_k}\left(d\phi+{\cal A}_k\right)^2,
\eeq
which is  topologically $\mathbb{CP}^2$ with orbifold singularities inherited from the spindle spanned by $(\alpha,\chi)$ ie $\mathbb{R}^4/\mathbb{Z}_{l_{k-1}}$ and $\mathbb{R}^4/\mathbb{Z}_{l_{k}}$ specifically, as well as a further orbifold singularity as $\sin\mu \to 0$, where $\mathbb{B}_4$  approaches a cone over $\mathbb{B}_3$ defined above. One can compute the Euler characteristic of this orbifold through the formula
\beq
\chi_E= \frac{1}{32\pi^2}\int_{\mathbb{B}_4}\left(R_{abcd}R^{abcd}-4 R_{ab}R^{ab}+R^2\right)\text{vol}(\mathbb{B}_4),
\eeq
which is a consequence of the Chern-Gauss-Bonnet theorem. Performing this integral is rather tedious but requires no special trick, the result is
\begin{equation}
\chi_E=\frac{l_kl_{k-1}+l_k(l_{k-1}-l_{k})+ l_{k-1}(l_{k-1}-l_{k})  }{l_{k-1}(l_{k-1}-l_k)l_k}=3- \left(1- \frac{1}{l_{k-1}}\right)- \left(1- \frac{1}{l_{k-1}-l_k}\right)- \left(1- \frac{1}{l_{k}}\right),   
\end{equation}
yielding the expected rational result, where 3 is the Euler characteristic on round $\mathbb{CP}^2$. We thus see that our manifold $\mathbb{B}_4$ is the weighted projective space $\mathbb{WCP}^2_{[l_{k-1},l_k,l_{k-1}-l_k]}$ which is a four dimensional analogue of a spindle. {Distinct examples of a restricted form of this orbifold, namely $\mathbb{WCP}^2_{[1,1,2]}$, have appeared before in \cite{Gauntlett:2004zh,Bianchi:2021uhn}. Here we have a generalisation that depends on two independent parameters\footnote{While an embedding of $\mathbb{WCP}^2_{[k_1,k_2,k_3]}$ (with $k_i$ truly independent) into supergravity probably exists, it is hard to see how it would be consistent with supersymmetry. For example,  $\mathbb{WCP}^2_{[1,1,1]}=\mathbb{CP}^2$, which such a solution would accommodate, certainly is not SUSY. We thank Dario Martelli for calling our attention to the paper \cite{Bianchi:2021uhn}}, to our knowledge this is the first time it has appeared in a solution of supergravity.} 

It is well known that reducing $\mathbb{R}^{1,5}\times \text{TN}_{M}$ on the Hopf fiber of the Taub-Nut space leads to the $d=7$ KK monopole geometry describing a stack of $M$ D6 branes in flat space. What we appear to have in \eqref{eq:interestingmonopole}  is the singularity associated to a $d=5$ KK monopole extended in AdS$_5$, that descends from the embedding of some conical Calabi-Yau 3-fold with orbifold singularities into $d=11$ via dimensional reduction. We give more details on this in appendix \ref{sec:KKdetials} where we show that the flat space analogue of the above singular behaviour can be realised by dimensional reduction of $d=11$ supergravity on an orbifold of $\mathbb{R}^{1,4}\times \mathbb{R}^{6}$. We can find the charge of this KK monopole by integrating $F_2$ at $\mu= \frac{\pi}{2}$, namely
\beq
-\frac{1}{2\pi}\int_{\mathbb{WCP}^1_{[l_{k-1},l_k]}}F_2= \frac{2N_k-N_{k+1}-N_{k-1}}{l_{k}l_{k-1}},\label{quantcond2}
\eeq
just as for the D6 branes in the previous deformation, only this time supersymmetry is not broken.

We now move to study some aspects of CFTs dual to our backgrounds.

\section{Comments on dual CFTs}\label{secCFT}
We start this section with generic comments on ${\cal N}=2$ super conformal field theories and their deformations. We propose that the CFT deformations we encountered in the (dual) description given by eq.(\ref{eqn:generalresult1}) represents marginal deformations.  We then analyse observables
like the central charge, showing that all the family of solutions obtained by deformations have the same holographic central charge. We comment on a mirror-like relation that our CFTs satisfy, and study spin-two fluctuations of our backgrounds. 
\subsection{General comments}

Conformal Field Theories (CFTs) play an important role among quantum field theories, as they allow for exact results, difficult to obtain for massive theories. In a given CFT, operators are classified as either irrelevant, marginal or relevant. Deformations control the RG-flow away from the fixed point. Given,
\begin{equation}
S= S_{CFT} + g \int d^dx~ {\cal O},\label{deformationCFT}
\end{equation}
with ${\cal O}$ a scalar primary field of dimension $\Delta$. The case for which $\Delta=d$ is specially interesting. The deformation becomes marginal in this case.

The dimension of any operator is often corrected by quantum effects (a beta-function for the coupling $g$ is induced). When the operator ${\cal O}$ is exactly marginal, the perturbation by such an operator gives place to a family of CFTs near the original fixed point. If two or more of such operators exist, one talks about a conformal manifold.
The existence of a conformal manifold requires the vanishing of the beta functions for all the couplings $g_i$ in eq.(\ref{deformationCFT}), $\beta(g_i)=0$. This is difficult to come by without the presence of SUSY or some other symmetry `protecting' the system from such corrections.

For the case of $d=4$ with ${\cal N}=1$ SUSY, Leigh and Strassler \cite{Leigh:1995ep} explained how the beta-functions of gauge and superpotential couplings are related, implying the existence of marginal operators and a conformal manifold.  A more powerful approach is presented in \cite{Green:2010da}. 

 In the context of AdS/CFT, conformal manifolds are mapped to AdS-vacua of supergravity theories. See for example \cite{Kol:2002zt}, \cite{Lunin:2005jy}.

Let us focus the attention on the non-SUSY ${\cal N}=0$ backgrounds.  The CFT dual to our family of backgrounds should admit a large $N$ expansion (we have also the parameter $P$, the length of the quiver). All the single trace operators with spin bigger than two must have very large dimension. CFTs with these characteristics were studied in \cite{Bashmakov:2017rko}.

One may wonder if such large $N$ CFT is still conformal after $\frac{1}{N}$ corrections are imposed, that is, if going beyond supergravity the isometries of AdS$_5$ (or those of S$^2\times$S$^1$) are still present. Using a bottom-up perspective, the SO(2,4) symmetry was shown to survive $\frac{1}{N}$ corrections in  \cite{Bashmakov:2017rko}.

Take the generic background  of eq.(\ref{eqn:generalresult1}), for the case $\zeta=0$. Consider its putative reduction to five dimensional gravity. The scalars in the AdS-bulk corresponding to marginal operators have mass $m^2=\Delta(\Delta-4)=0$. Hence, the non-normalisable mode of those scalars is dual to the coupling $g$ in the perturbed CFT in eq.\eqref{deformationCFT}. The conformal manifold is associated with the moduli space of AdS$_5$ vacua in the reduced theory. 

{We emphasise that the above arguments are heuristic. Finding exactly marginal operators in situations without SUSY is not a well-understood problem. At best, we state that if our family of backgrounds is stable (proving this requires a more dedicated study than the one we aim at here), the supergravity solutions would be the best indicators of the existence of such exactly marginal deformations. }

\subsection{The dual to our backgrounds}
Let us now go into more detail for the marginal deformations generically represented by the background in eq.  (\ref{eqn:generalresult1}). We focus on the ${\cal N}=1$ and $U(1)_R$ preserving case (for $\xi=-\zeta$) and in the $\text{SU}(2)$ preserving ${\cal N}=0$ case (with $\zeta=0$).

Using the representation theory of the superconformal algebra, see the paper \cite{Cordova:2016emh}, the work \cite{Xie:2019aft} analysed soft-SUSY breaking of 4d ${\cal N}=2$ SCFTs.

Four dimensional ${\cal N}=2$ SCFTs have global symmetries (bosonic part)
 given by,
 \begin{equation}
 \text{SO}(2,4) \times \text{SU}(2)_R\times \text{U}(1)_R \times \text{G}_F.
 \nonumber
 \end{equation} 
 The first factor is the conformal group in four dimensions
and corresponds with the isometries of AdS$_5$. The SU(2)$_R\times$U(1)$_R$  R-symmetry is associated with the S$^2\times$S$^1$ part of the geometry in \eqref{eq:N=2}. The factor $G_F$ represents other global symmetries, like flavour symmetries which are realised on the world-volume of D6 branes.

A highest weight state is labelled as $|\Delta, R, r, j_1,j_2>$. Here $\Delta$ represents the scaling dimension, $R$ is the charge under SU(2)$_R$, while $r$ is the charge under U(1)$_R$.  the values $(j_1,j_2)$ are the left and right spin, when we consider SO(1,3)$\sim$SO(4)$\sim$ SU(2)$\times$ SU(2). The short representations have been classified in \cite{Dolan:2002zh}, \cite{Cordova:2016xhm}.

We are specially interested in Coulomb branch operators, denoted by ${\cal E}_{(r,0,0)}$. These operators have component fields: $A$ (a scalar), $\Psi^i$ (a spinor in the fundamental of SU(2)$_R$), $B^{(ij)}$ (scalars in the adjoint of SU(2)$_R$), $F_{\alpha\beta}$ (an anti-self-dual two form), $\Lambda^i$ (a spinor in the fundamental of $\text{SU}(2)_R$) and $C$ (a scalar). We are interested in the $[\Delta, \text{SU}(2)_R, \text{U}(1)_R ]$ values of the scalar components of the multiplets, as these can be used to deform the theory. These values are  \cite{Xie:2019aft}
\begin{equation}
A=[r,0,r],~~ B^{(ij)}=[r+1,1,r-1],~~C=[r+1,0, r-2].
\end{equation}
We can consider deformations in eq.(\ref{deformationCFT}) with the form $g_i \int d^4 x ~{\cal O}_i$. For the operators
 \begin{equation}
 {\cal O}_1= B^{(12)}+ cc,~~~
  {\cal O}_2= B^{(11)}+ cc,~~~ {\cal O}_3= B^{(22)}+ cc.~~~
 \end{equation}
 In  \cite{Xie:2019aft}, it is shown that these deformations have dimension $\Delta=r+1$. Choosing $r=3$ we have marginal operators. For the case of the deformation with the operator ${\cal O}_1$, we have a preserved SU(2) global symmetry, inherited from the R-symmetry, and SUSY is completely broken to ${\cal N}=0$. We associate this deformation with the line of CFTs described by the parameter $\xi$ (with $\zeta=0$). Similarly, the operators ${\cal O}_2$ and ${\cal O}_3$
 preserve ${\cal N}=1$ SUSY and the associated R-symmetry is  $\text{U}(1)= \frac{2}{3} \left( \text{U}(1)_R \pm 2 I_3\right)$.
 In fact, the scaling dimension of these operators is $\Delta=4$. The amount of SUSY and global symmetry suggests that these deformations with ${\cal O}_2, {\cal O}_3$ correspond with the branch $\xi=-\zeta$ in eq. (\ref{eqn:generalresult1}). The quantisation conditions found in eqs.(\ref{quantcond1}) and (\ref{quantcond2}), associated with the presence of spindles, suggest that these deformations are non-Lagrangian.

{The arguments above are not air-tight, particularly those in the SUSY breaking family $\zeta=0$. The study of exactly marginal deformations requires a more careful analysis than the one offered here. To ascertain the relevance of our $\zeta=0$ family of backgrounds, a more dedicated stability analysis should be performed. This is outside the scope of this work.}

\subsection{Central Charges}
The c-function is a quantity defined at fixed points of the renormalisation group flow.
In two dimensions, it was proven by Zamolodchikov \cite{Zamolodchikov:1986gt}
 that (with reasonable assumptions) under an RG-flow $\frac{d c(t)}{dt} \leq 0$. Here $t=-\log\left( \frac{\mu}{\Lambda} \right)$ and $c(t)$ is the central charge, the coefficient appearing in the correlator of two energy momentum tensors or in the trace anomaly. Thanks to this, the $c$-function can be used at the fixed points as a measure of the number of degrees of freedom of the CFT.
 
 In the case of four dimensional supersymmetric field theories, two possible central functions appear in the correlator of energy momentum tensors or
in the trace anomaly. These quantities are called $a$ and $c$. It was shown by Komargodsky and Schwimmer \cite{Komargodski:2011vj} that (given some reasonable assumptions), the quantity $a$ is monotonically decreasing towards the IR $\frac{d a(t)}{dt} \leq 0$. In particular, it can be used as a measure of the number of degrees of freedom.

In the special case of conformal long linear quivers with ${\cal N}=2$ SUSY that we consider in this work, with $N_v$ vector multiplets and $N_h$ hypermultiplets, it can be shown (see for example \cite{Shapere:2008zf})
that,
\begin{equation}
a=\frac{5 N_v + N_h}{24},~~~~c=\frac{2 N_v + N_h}{12}.\label{centralsN=2}
\end{equation}
Holographically (in the supergravity approximaton) it is shown that $a=c$, see for example \cite{Henningson:1998gx}. The corrections to this relation are suppressed by the numbers $\frac{1}{N}$ and $\frac{1}{P}$, being $N$ a generic gauge group rank (considered to be large) and $P$ the (large) length of the linear quiver.

We define below a string theory quantity that in the case of ${\cal N}=2$ long linear quivers has been shown to match precisely with the results for $a$ and $c$ computed using localisation and matrix model techniques \cite{Nunez:2023loo}. This quantity is nothing but a generalisation to more generic backgrounds of well-known formulas. In other words, the inverse of the lower dimensional Newton constant is calculated and this is associated with the free energy and with the number of degrees of freedom. We refer to this quantity as holographic central charge.

\subsubsection{Holographic Central Charge}

We now turn to calculate the Holographic Central charge.  On the CFT side, it is one of the key characteristic quantities, the Free Energy of the CFT on S$^4$ (counting the number of degrees of freedom). On the Supergravity side, it measures a weighted effective volume of the internal manifold. See \cite{Bea:2015fja,Macpherson:2014eza,Henningson:1998gx,Klebanov:2007ws} for further details.
To calculate this internal volume, we follow the methodology outlined in \cite{Macpherson:2014eza}, in which, given a metric of the form
\beq
ds^2=\alpha(\rho,\vv{\theta})\Big(dx_{1,d}^2+\beta(\rho)d\rho^2\Big)+g_{ij}(\rho,\vv{\theta})d\theta^id\theta^j,   
\eeq
we define
\begin{equation}
\begin{gathered}
V_{int}=\int d\vv{\theta}\sqrt{\text{det}[g_{ij}]e^{-4\Phi}\alpha^d},~~~~~~~~~~~~~~~~~~~~~~H=V_{int}^2,
\end{gathered}
\end{equation}
with the corresponding Holographic Central Charge, $c_{hol}$, given by 
\beq
c_{hol}=\frac{d^d}{G_N}\beta^{d/2}\frac{H^{\frac{2d+1}{2}}}{(H')^d},
\eeq
where $G_N=8\pi^6 \alpha'^4g_s^2=8\pi^6$ (in the units $\alpha'=g_s=1$).
For the backgrounds presented in eq.(\ref{eqn:generalresult1}), we have
\beq
\begin{gathered}
d=3,~~~~~~~~~~~~~\alpha=4\rho^2e^{\frac{2}{3}\Phi}f_1,~~~~~~~~~~~\beta=\frac{1}{\rho^4},\\
g_{ij}(\rho,\vv{\theta})d\theta^id\theta^j =  e^{\frac{2}{3}\Phi}f_1 \bigg[f_2 ds^2(\text{S}^2)+f_4 (d\sigma^2+d\eta^2)\bigg]+e^{-\frac{2}{3}\Phi} f_1^2f_5 f_3  d\chi^2.
\end{gathered}
\eeq
After lengthy algebra that is described in detail in Appendix \ref{sec:HCC}, one arrives at the  result, 
 \begin{equation}\label{eqn:HCC}
\begin{aligned}
c_{hol}=\frac{1}{4\pi}\int_0^P \mathcal{R}(\eta)^2d\eta =\frac{1}{8\pi}\sum_{k=1}^\infty P \, \mathcal{R}_k^2,
\end{aligned}
\end{equation}
where all dependence on the dilaton drops out neatly, meaning the Holographic Central Charge is the same for all backgrounds presented throughout this paper. 
Indeed, eq.\eqref{eqn:HCC} matches the $\mathcal{N}=2$ result of \cite{Nunez:2019gbg} (up to appropriate conversion of notation). The  interested reader might want to calculate the result of eq.(\ref{eqn:HCC}) for a given Rank function and compare it with the result of eq.(\ref{centralsN=2}). Various examples along these lines are worked out in  \cite{Nunez:2019gbg},  \cite{Nunez:2023loo}, \cite{Lozano:2016kum}.

The fact that the central charge does not change when considering the ${\cal N}=(2,1,0)$ backgrounds, indicates that the CFT dual to these different backgrounds are related by marginal deformations. This is in nice coincidence with the presence of parameters $(\xi,\zeta)$ that control these deformations, as proposed above. \\
~\\
Let us now  study another aspect of these ${\cal N}=2$, ${\cal N}=1$ and ${\cal N}=0$ CFTs.


\subsection{A mirror-like relation}
Consider a generic rank function 
\begin{equation}\label{eqn:Rkoriginal1}
\mathcal{R}(\eta) =
    \begin{cases}
     ~~~~~~~~~~~~~~~ N_1\eta & \eta\in[0,1]\\
      N_k+(N_{k+1}-N_k)(\eta-k) & \eta\in[k,k+1]\\
   ~~~~~~~~~   N_{P-1}(P-\eta)  & \eta\in[P-1,P].
    \end{cases}     
\end{equation}
The total number of flavours in the associated linear quiver is given by $F ={\cal R}'(0) -{\cal R}'(P)= N_1+ N_{P-1}$. The length of the quiver is $(P-1)$, that is the number of gauge nodes.

 Let us assume in this section that both $\frac{F}{P}$ and $\frac{P}{F}N_j$ are integer numbers. 
Let us now define a second rank function ${\hat{ \cal R}}(\hat{\eta})$ to be,
\begin{equation}\label{eqn:Rkoriginal2}
\hat{\mathcal{R} }(\hat{\eta}) =
    \begin{cases}
     ~~~~~~~~~~~~~~~ \hat{N}_1\hat{\eta} & \hat{\eta} \in [0,\frac{F}{P}]\\
      \hat{N}_k+(\hat{N}_{k+1}-\hat{N}_k) ( \hat{\eta}-k) & \hat{\eta} \in  [k \frac{F}{P},(k+1)\frac{F}{P} ] \\
   ~~~~~~~~~   \hat{N}_{F-1}(F-\hat{\eta})  & \hat{\eta}\in [F(1-\frac{1}{P}),F],
    \end{cases}     
\end{equation}
where $\hat{N}_j= \frac{P}{F} N_j$.
The total number of flavours is $P= \hat{ {\cal R} }'(0) -\hat{ \cal{R}  }'(F)= \hat{N}_1-\hat{N}_{F-1}$. The length of the quiver is $(F-1)$.

One can easily show that the Fourier coefficients of ${\cal R}(\eta)$ and $\hat{ {\cal R} } (\hat{\eta}  ) $ are identical. In other words,
\begin{equation}
{\cal R}_n= \frac{1}{P}\int_{-P}^P { \cal R}(\eta) \sin\left( \frac{n \pi \eta}{P}\right) d\eta= \frac{1}{F}\int_{-F}^F \hat{ \cal R}(\hat{\eta}) \sin\left( \frac{n \pi \hat{\eta}}{F}\right) d\hat{\eta}= \hat{\cal R}_n.
\end{equation}
Under these conditions, it was shown in  \cite{Nunez:2023loo} that the two linear quivers have the same `density of free energy' or `holographic central charge per unit length'. That is,
\begin{equation}
\frac{c_{hol}}{P}= \frac{1}{8\pi} \sum_{k=1}^\infty {\cal R}_k^2=  \frac{1}{8\pi} \sum_{k=1}^\infty \hat{ {\cal R}}_k^2= \frac{\hat{c}_{hol}}{F}.
\end{equation}
This relation between two different quivers and their free energy per unit length, first observed in  \cite{Nunez:2023loo} is an extension of mirror symmetry to the four dimensional case, with either ${\cal N}=(2,1,0)$. 

Let us now study a special type of excitations in our CFTs.

\subsection{Spin 2 fluctations}\label{spintwofluct}
In this section we study particular excitations in our family of backgrounds. These are excitations of the metric, along the directions of AdS$_5$. This simple fluctuation is consistent and can be associated with states of spin two in the CFT. This kind of excitations  were object of study in  SCFTs in different dimensions. A precursor to these studies is \cite{Csaki:2000fc}. In this work we rely on the results of  \cite{Bachas:2011xa,Chen:2019ydk,Itsios:2019yzp,Lima:2023ggy}. 
In fact, following   \cite{Bachas:2011xa,Chen:2019ydk,Itsios:2019yzp}, for solutions of the following geometry (in this section we work in {\it Einstein frame}), where
\begin{equation}
ds_E^2= e^{2A_E}ds^2(\text{AdS}_5) + ds^2(\mathcal{M}_5),
\end{equation}
with spin-two fluctuations along only the AdS$_5$ part of the metric in Einstein Frame, where  the ten coordinates are labelled $X^M =(x^\mu,y^a)$,
\begin{equation}
\begin{aligned}
ds_E^2 &= e^{2A_E}\bigg[\Big(\tilde{g}_{\mu\nu}(x) +h_{\mu\nu}(x,y)\Big)dx^\mu dx^\nu + \tilde{g}_{ab}(y)dy^a dy^b\bigg],\label{genericglue}
\end{aligned}
\end{equation}
with the condition that the fluctuation $h_{\mu\nu}$ written in terms of a tensor that is transverse and traceless as,
\begin{equation}
h_{\mu\nu}(x,y) = h_{\mu\nu}^{[tt]}(x)\mathcal{F}(y),~~~~~~~~~~~~\tilde{\nabla}^\mu  h_{\mu\nu}^{[tt]}=0,~~~~~~~~~~~~~\tilde{g}^{\mu\nu}h_{\mu\nu}^{[tt]}=0.
\end{equation}
As is discussed in  \cite{Bachas:2011xa,Itsios:2019yzp}, the fluctuation of Maxwell equations and Dilaton equation are satisfied trivially. However, the Einstein equations lead to the following condition,
\begin{equation}\label{eqn:Spin2-1}
\begin{aligned}
0 &= \tilde{\nabla}^\sigma \tilde{\nabla}_\sigma h_{\mu\nu} +2 h_{\mu\nu} + \tilde{\nabla}^a \tilde{\nabla}_a h_{\mu\nu} + 8 \tilde{\nabla}^a A \tilde{\nabla}_a h_{\mu\nu}\\
&= \tilde{\nabla}^\sigma \tilde{\nabla}_\sigma h_{\mu\nu} +2 h_{\mu\nu} + e^{-8A_E} \tilde{\nabla}^a \Big[e^{8A_E} \tilde{\nabla}_a h_{\mu\nu}\Big] \\
&:= \tilde{\nabla}^\sigma \tilde{\nabla}_\sigma h_{\mu\nu} +2 h_{\mu\nu} + \mathcal{L}(h_{\mu\nu}),
\end{aligned}
\end{equation}
noting that $h_{\mu\nu}$ acts like a scalar for $\tilde{\nabla}_a$. This corresponds to the equation of motion for a graviton propagating on AdS$_5$, given by the Pauli-Fierz equation
\begin{equation}
 \tilde{\nabla}^\sigma \tilde{\nabla}_\sigma h_{\mu\nu} =(M^2-2)h_{\mu\nu},
\end{equation}
where $M$ is the graviton mass, meaning 
\begin{equation}
\begin{aligned}
 \mathcal{L}(h_{\mu\nu})=- M^2 h_{\mu\nu}. \label{gravitonmass}
\end{aligned}
\end{equation}
For some scalar fluctuation $\mathcal{F}$, we have comparing with (\ref{eqn:Spin2-1})-(\ref{gravitonmass}),
\begin{equation}\label{eqn:Leq}
\mathcal{L}(\mathcal{F}) =\frac{e^{-8A_E}}{\sqrt{\tilde{g}_{\mathcal{M}_5}}}\partial_a \Big(e^{8A_E}\sqrt{\tilde{g}_{\mathcal{M}_5}} \tilde{g}^{ab} \partial_b \mathcal{F}\Big)  =\frac{1}{\sqrt{\tilde{g}_{\mathcal{M}_5}}}\partial_a \Big(\sqrt{\tilde{g}_{\mathcal{M}_5}} \tilde{g}^{ab} \partial_b \mathcal{F}\Big) + 8 \tilde{g}^{ab}  \partial_a A \,\partial_b \mathcal{F}.
\end{equation}
In what follows we study the equation of motion for the fluctuation in the metrics analysed in this work.

 \subsubsection*{\textbf{$\mathcal{N}=0$ Reduction}}
Using the form of the  SU$(2)$ preserving $\mathcal{N}=0$ metrics presented in eq.(\ref{N=0metric}), after moving to Einstein Frame, one finds
\begin{equation}
ds_{E}^2=4(f_1^9 f_5 \Delta)^{\frac{1}{8}}\bigg[ds^2(\text{AdS}_5))+\frac{1}{4}\bigg(f_2 ds^2(\text{S}^2)+f_4(d\sigma^2+d\eta^2)+\frac{f_3 }{\Delta} d\chi^2\bigg)\bigg].
\end{equation} 
    Using eq.(\ref{genericglue}), this gives
    \begin{equation}
    e^{2A_E} = 4 (f_1^9 f_5 \Delta)^{\frac{1}{8}} ,~~~~~~~~\tilde{g}_{\mathcal{M}_5}=\frac{1}{4^5\Delta}f_2^2 f_4^2 f_3 \sin^2\theta,~~~~~~~~e^{8A_E}\sqrt{\tilde{g}_{\mathcal{M}_5}} = {2^3}f_1^{\frac{9}{2}}f_3^{\frac{1}{2}}f_5^{\frac{1}{2}}f_2f_4 \sin\theta.
    \end{equation}
    Using this in \eqref{eqn:Leq} and the definitions of $f_i$ in \eqref{eq:thefs} one finds,
\begin{align}
&\frac{2\left((2\dot{V}-\ddot{V})V''+(\dot{V}')^2\right)}{V''\dot{V}}\nabla_{\text{S}^2}^2 \mathcal{F}  +\frac{ 2\dot{V}- \ddot{V}}{\sigma^2 V''}\Delta\partial_\chi^2  \mathcal{F}\nn\\[2mm]
&+ \frac{2}{V''\dot{V}\sigma }\bigg[\partial_\eta\Big( \sigma \dot{V}^2   \partial_\eta\Big)  +\partial_\sigma\Big( \sigma \dot{V}^2  \partial_\sigma\Big) \bigg] \mathcal{F} +M^2 \mathcal{F}  =0.
\end{align}
which only differs from the ${\cal N}=2$ result by $\Delta$, which of course goes to 1 when $\xi=0$ and where the Laplacian on the two sphere is 
\beq
\nabla_{\text{S}^2}^2\equiv \frac{1}{\sin\theta}\partial_\theta (\sin\theta \partial_\theta)+\frac{1}{\sin^2\theta}\partial_\phi^2.
\eeq
Universal spin-two modes in the GM background were already considered in \cite{Chen:2019ydk}, following the procedure there we expand the mass eigenfunction as
\begin{equation}
\mathcal{F} = \sum_{lmn}\phi_{lmn}Y_{lm}e^{in\chi},~~~~\nabla_{\text{S}^2}^2Y_{lm} = -l(l+1) Y_{lm},\label{fredef}
\end{equation}
where $l,m,n\in\mathbb{Z}$ and $l\geq1$ and were we stress that $\mathcal{F}$, and so also $\phi_{lmn}$ are complex in general. This leads to 
\begin{align}
&-l(l+1)\frac{2\left((2\dot{V}-\ddot{V})V''+(\dot{V}')^2\right)}{V''\dot{V}} \mathcal{F}  -n^2\frac{ 2\dot{V}- \ddot{V}}{\sigma^2 V''}\Delta   \mathcal{F}\nn\\[2mm]
&+ \frac{2}{V''\dot{V}\sigma }\bigg[\partial_\eta\Big( \sigma \dot{V}^2   \partial_\eta\Big)  +\partial_\sigma\Big( \sigma \dot{V}^2  \partial_\sigma\Big) \bigg] \mathcal{F} +M^2 \mathcal{F}  =0.\label{eq:lnformneq0}
\end{align}
 Note that this is a version of the Sturm-Liouville problem and  fits into the `universal form'  given in \cite{Lima:2023ggy}, one should identify 
\begin{eqnarray}
& &\partial_a(\bar{p}\partial^a\psi)+\bar{q}\psi =-\bar{m}^2 \bar{w} \psi, ~~~~\text{with} ~~\psi=\phi_{\hat{l}\hat{m}\hat{n}} ~~~\text{and}  \label{eqn:UF}\\
& & \bar{p}=\sigma \dot{V}^2,~~~~~\bar{w}=\frac{1}{2}\sigma V'' \dot{V}, ~~~~~\bar{m}^2=M^2,~~~ \bar{q}= - \sigma \dot{V}V'' \left( \frac{\tilde{\Delta} \hat{l}(\hat{l}+1)}{\dot{V} V''} +\frac{\hat{n}^2 \Lambda \Delta}{2\sigma^2} \right).\nonumber
\end{eqnarray}
In \cite{Chen:2019ydk}, the ${\cal N}=2$ analogue of \eqref{eq:lnformneq0} is mapped to a more useful form by redefining $\phi_{lmn}$, here we do something similar, namely take
\beq
\phi_{lmn}= e^{n \xi V'}\sigma^n (\dot{V})^l\tilde{\phi}_{lmn},~~~~ M^2=\mu^2+(2l+n)(2l+n+4),
\eeq
upon which \eqref{eq:lnformneq0} becomes
\beq
\sigma^n \dot{V}^le^{n\xi V'}\left(\frac{2 e^{-2n \xi V'}}{\sigma^{2n-1}\dot{V}^{2l+1}\ddot{V}}\partial_a\left(\sigma^{2n+1}\dot{V}^{2+2l}e^{2n\xi V'}\partial_a\tilde{\phi}_{lmn}\right)-\mu^2\right)=0,\label{eq:spin2finalneq0}
\eeq
where $a\in(\eta,\sigma)$. This is solved by $\tilde{\phi}_{lmn}=$ constant and $\mu=0$ for all $n$, this leads to the universal solution
\beq
\phi^r_{lmn}= e^{n \xi V'}\sigma^n (\dot{V})^l\phi_0,~~~~ M^2=-4+(2+2l+n)^2,
\eeq
where $\phi_0=$ constant. This solution remains finite at all points on the Riemann surface for the $V$ defined in section \ref{sec:nequals2solution}, just as they were shown to be for the ${\cal N}=2$ case in \cite{Chen:2019ydk}.

We can derive an integration measure to define the norm of fluctuations by computing the quadratic fluctuation of  $S= \frac{1}{2\kappa_{10}^2}\int\sqrt{-\det g} R dx^{10}$, we find in general
\beq
\delta^{(2)} S= \frac{1}{2\kappa_{10}^2}\int dx^5 dy^5 e^{3A}\sqrt{-\det g_{\text{AdS}_5}}\sqrt{\det {\cal M}_5}
h_{\mu\nu}\left(\nabla^2_{\text{AdS}_5}+2-M^2\right)h^{\mu\nu}.
\eeq
We thus see that an appropriate integration measure is $e^{3A}\sqrt{\det {\cal M}_5}$ when integrating over $y$. We can use this to derive a bound on  $M^2$. Starting from \eqref{eq:spin2finalneq0} and contracting with $\overline{\phi}_{lmn}$ and then integrating with respect to the above measure we find
\beq
-\int d\eta d\sigma\bar{\tilde{\phi}}_{lmn}\left(\partial_a\left(\sigma^{2n+1}\dot{V}^{2+2l}e^{2n\xi V'}\partial_a\tilde{\phi}_{lmn}\right)+\frac{1}{2}\mu^2 e^{2n\xi V'}\sigma^{2n+1}\dot{V}^{2l+1}V'' \tilde{\phi}_{lmn}\right)=0
\eeq
Integration by parts, assuming no boundary contributions, which is the case if the fluctuation is finite everywhere on the Riemann surface, as a regular fluctuation must be, then yields
\beq
\int d\eta d\sigma\left(\sigma^{2n+1}\dot{V}^{2+2l}e^{2n\xi V'}|\partial_a\tilde{\phi}_{lmn}|^2-\frac{1}{2}\mu^2 e^{2n\xi V'}\sigma^{2n+1}\dot{V}^{2l+1}V'' |\tilde{\phi}_{lmn}|^2\right)=0 \label{diarma}
\eeq
where the first term is positive definite and the second term is negative definite. We see that the minimal value  $\mu^2$ can take is $\mu^2=0$, which happens for  $\tilde{\phi}_{lmn}$ constant. We conclude that
\beq
M^2\geq (2l+n)(2l+n+4),\label{zaza}
\eeq
just as was the case for ${\cal N}=2$. Note that as a bound on the scaling dimension of operators this becomes $\Delta \geq 4+2l+n$.

{Finally, let us emphasise that the result for the ${\cal N}=0$ family of backgrounds (parametrised by $\xi$) is a deformation of the 
result of the ${\cal N}=2$ backgrounds. See eq.(\ref{diarma}). This suggest that the positive masses obtained in (\ref{zaza}) are an indication of stability of our SUSY-breaking family of backgrounds.}

 \subsubsection*{\textbf{$\mathcal{N}=1$ Reduction}}
Using the form of the  $\mathcal{N}=1$ metrics presented in eq.\eqref{eq:neq1sol}, after moving to Einstein Frame,
    \begin{align}
        ds_{E}^2=4(f_1^9f_5\Xi)^{\frac{1}{8}}\bigg[ds^2(\text{AdS}_5)+h_{\mu\nu} dx^\mu dx^\nu+\frac{1}{4}\bigg(f_4(d\sigma^2+d\eta^2)+ds^2(M_3) \bigg)\bigg],
    \end{align}
 with $ds(M_3)^2$ given in eq.\eqref{eqn:M3}. Using eq.\eqref{genericglue} gives
    \begin{equation}
    e^{2A_E} =4(f_1^9f_5\Xi)^{\frac{1}{8}},~~~~~~~~\tilde{g}_{\mathcal{M}_5}=\frac{1}{4^5\Xi}f_2^2 f_4^2  f_3 \sin^2\theta ,~~~~e^{8A_E}\sqrt{\tilde{g}_{\mathcal{M}_5}} = 2^3 f_1^{\frac{9}{2}}f_3^{\frac{1}{2}}f_5^{\frac{1}{2}}f_2f_4\sin\theta.~~~~
    \end{equation}
This leads to the differential equation
\begin{align}
&\frac{2\left((2\dot{V}-\ddot{V})V''+(\dot{V}')^2\right)}{V''\dot{V}}\nabla_{\text{S}^2}^2 \mathcal{F}  +\frac{ 2\dot{V}- \ddot{V}}{\sigma^2 V''}\Delta\partial_\chi^2  \mathcal{F}\nn\\[2mm]
&+\left(4\xi^2\left(\frac{1}{f_5}+\frac{f_6^2}{f_3}\right)\partial_{\phi}^2+8 \xi \frac{\xi f_3 +f_5 f_6(1+\xi f_6)}{f_3f_5}\partial_{\chi}\partial_{\phi}\right){\cal F}\nn\\[2mm]
&+ \frac{2}{V''\dot{V}\sigma }\bigg[\partial_\eta\Big( \sigma \dot{V}^2   \partial_\eta\Big)  +\partial_\sigma\Big( \sigma \dot{V}^2  \partial_\sigma\Big) \bigg] \mathcal{F} +M^2 \mathcal{F}  =0.
\end{align}
Despite appearances, this actually behaves remarkably similar to the equation defining the mass of spin 2 fluctuations in the ${\cal N}=0$ deformation.  We once more redefine ${\cal F}$ in terms of  \eqref{fredef}, and then this time further redefine 
\beq
\phi_{lmn}= e^{(n-m) \xi V'}\sigma^n (\dot{V})^l\tilde{\phi}_{lmn},~~~~ M^2=\mu^2+(2l+n)(2l+n+4).
\eeq
Again making use of $\nabla_{\text{S}^2}^2Y_{lm} = -l(l+1) Y_{lm}$ and that $\partial_{\phi}Y_{lm}=im Y_{lm}$ on finds that the resulting PDE for $\tilde{\phi}_{lmn}$ takes an almost identical form to \eqref{eq:spin2finalneq0}, except with $e^{n \xi V'}$ replaced with $e^{(n-m) \xi V'}$ everywhere it appears. The remaining arguments of the previous section then go through the same: A universal regular fluctuation, valid when $\mu^2=0$ is given by
\beq
\phi^r_{lmn}= e^{(n-m) \xi V'}\sigma^n (\dot{V})^l\phi_0,~~~~ M^2=-4+(2+2l+n)^2
\eeq
where $\phi_0$ is constant. This solution again saturates the bound on the mass on can derive, which is again
\beq
M^2\geq (2l+n)(2l+n+4),
\eeq
once more matching the ${\cal N}=2$ result.

 \section{Conclusions and future directions}\label{concl}
Let us start with a brief summary of the new things presented in this work.
\begin{itemize}
\item{After  reviewing the electrostatic formulation of holographic duals to ${\cal N}=2$ four dimensional linear quiver SCFTs and a careful study of the quantised charges, we wrote the SU$(2)$ G-structure in terms of two-forms $j,\omega$ and one forms $u,v$. We wrote the pure spinors and in terms of these found calibrated branes in these holographic duals.} 
\item{Relying on an SL$(3,\mathds{R})$ transformation in eleven dimensions, we constructed a new infinite family of  solutions. Each member of the family is labelled by a potential function (one for each parent linear quiver) $V(\sigma,\eta)$, and two numbers $(\xi,\zeta)$. We studied SUSY preservation for different values of $(\xi,\zeta)$, wrote the G-structure when possible and studied quantised charges.  Interestingly the charges reveal that the space orthogonal to the branes is a spindle in general. This may translate in the non-Lagrangian character of the dual SCFTs. The Figure \ref{figure1xx} indicates that for generic values of $(\xi,\zeta)$ the backgrounds break SUSY completely (${\cal N}=0$). Interestingly, there is a family of ${\cal N}=0$ backgrounds with an SU$(2)$ global symmetry, inherited from the parent R-symmetry. There is also a family of ${\cal N}=1$ backgrounds with its corresponding U$(1)_R$ symmetry. }
\item{We studied some sample observables in the dual SCFTs. In particular, we proposed which operators may be marginally deforming the ${\cal N}=2$ SCFTs. We calculated the holographic central charge, a quantity associated with the $a$ central charge. In the holographic limit $a=c$. We showed the independence of the holographic central charge on the parameters $(\xi,\zeta)$. This being in agreement with the proposed marginally deformed CFT. We proposed a mirror-like relation between pairs of quivers. This relation is valid for all members of the family. We  calculated spin-two fluctuations in the CFT and wrote  equations ruling their dynamics. Universal solutions to these equations are written, together with the spectrum of dimensions. A lower bound is proven for the spectrum of dimensions.}
\end{itemize}
For the future, this work opens some interesting
topics to be studied.
Here we write some things that would be interesting to work in detail.
\begin{itemize}
\item It would be interesting to see whether the $d=4$ weighted projective space we find during our analysis of the ${\cal N}=1$ preserving deformation can appear  in more standard wrapped brane scenarios, \textit{i.e.} compactifications of AdS$_{n+4}$ to AdS$_{n}$   for $n=2,3$.
\item{Use our thrice-infinite family of new solutions to holographically calculate observables. For example, the calculations in \cite{Nunez:2018qcj}-\cite{Roychowdhury:2021eas} are suitable to be done.}
\item{A better understanding of the operators that trigger the marginal deformation is desirable.}
\item{It would be interesting to repeat the analysis in this paper in two related systems (with qualitatively different) holographic field theory. One is the Lin-Maldacena family of solutions dual to different vacua of the BMN matrix model \cite{Lin:2005nh}, see the papers \cite{vanAnders:2007ky}-\cite{Lozano:2017ole} for different elaborations on the backgrounds. The other is the system describing a 4d defect inside the $(0,2)$ 6d SCFT. See for example \cite{Capuozzo:2023fll}.
}
\item{It would be nice to study the SUSY probe-dynamics we encountered in the ${\cal N}=2$ and ${\cal N}=1$ systems.}
\item{It would be interesting to place our ${\cal N}=1$ SUSY family of solutions in the context of the work \cite{Ashmore:2021mao}. Possibly, the ${\cal N}=0$ family of backgrounds can be understood within the formalism developed in \cite{Menet:2023rnt}.}
\item{Recently,  studies on the stability (or not) of non-SUSY backgrounds have been refined thanks to the application of techniques of Exceptional field theory, see for example \cite{Malek:2020yue}-\cite{Eloy:2023acy}. It would be interesting to apply this technology to our family of ${\cal N}=0$ backgrounds, for any values of the $(\xi,\zeta)$ parameters. Stability studies along the lines of the paper \cite{Apruzzi:2021nle} seem pertinent for our backgrounds.}

\end{itemize}
We hope to soon report on these and other interesting topics.

\section*{Acknowledgements}
We thank Christopher Couzens for several enlightening discussions. The work of NM is supported by the Ram\'on y Cajal fellowship RYC2021-033794-I, and by grants from the Spanish government MCIU-22-PID2021-123021NB-I00 and principality of Asturias SV-PA-21-AYUD/2021/52177.
P.M and C.N are supported by grants ST/X000648/1 and ST/T000813/1.

\appendix

\section{Transformation from LLM to GM}\label{sec:Transformation}

We now present a step-by-step transformation from the LLM to the GM background via the `B$\ddot{\text{a}}$cklund' transformation, demonstrating explicitly \eqref{eqn:pickup} (as described in \cite{Bah:2019jts,Bah:2022yjf,Couzens:2022yjl})
\begin{equation}\label{eqn:pickup}
\bar{\chi}\rightarrow\chi+ \beta,~~~~~~\bar{\beta}\rightarrow -\beta.
\end{equation}
We begin with the LLM background \cite{Lin:2004nb}\cite{Macpherson:2016xwk}\cite{Bah:2019jts} 
\begin{align}
ds^2&= \kappa^{\frac{2}{3}}e^{2\lambda}\bigg[4ds^2(\text{AdS}_5)+y^2 e^{-6\lambda} ds^2(\text{S}^2)+\frac{4}{1-y\partial_yD}\Big(d\bar{\chi}-\frac{r}{2}\partial_rD d\bar{\beta}\Big)^2-\frac{\partial_yD}{y}\Big[dy^2+e^D (dr^2+r^2d\bar{\beta}^2)\Big]\bigg],\nn\\[2mm]
G_4&= \kappa \bigg[-d(2y^3 e^{-6\lambda}) \wedge d\bar{\chi} +  \bigg(d\bigg[e^{-6\lambda}\frac{y^2}{\partial_yD}r\partial_rD\bigg] - \partial_y(e^D)r\,dr + r\partial_rD\,dy\bigg) \wedge d\bar{\beta} \bigg]\wedge \text{vol}(\text{S}^2),\label{eqn:LLM}
\end{align}
with $D(r,y)$ satisfying
\begin{equation}
\frac{1}{r}\partial_r(r \partial_r D) +\partial_y^2e^D=0,~~~~e^{-6\lambda} = \frac{-\partial_yD}{y(1-y\partial_yD)}.
\end{equation}
By direct comparison with the GM case \eqref{eqn:GM}, re-written below for clarity
\begin{align}
ds^2&=f_1 \Bigg[4ds^2(\text{AdS}_5) +f_2 ds^2(\text{S}^2)+f_3 d\chi^2 +f_4 \big(d\sigma^2 +d\eta^2\big)+f_5 \Big(d\beta +f_6 d\chi\Big)^2\Bigg],\\[2mm]
A_3&=\Big(f_7 d\chi +f_8 d\beta \Big)\wedge \text{vol}(\text{S}^2),
\end{align}
it is immediately clear that 
\begin{equation}
\kappa^{\frac{2}{3}}e^{2\lambda} = f_1,~~~~y^2 e^{-6\lambda} =f_2.
\end{equation}
After the following `B$\ddot{\text{a}}$cklund' transformation
\begin{equation}\label{eqn:Backlund}
r^2e^D=\sigma^2,~~~~y=\dot{V},~~~~\text{log}(r)=V',
\end{equation}
with the given relation for $\lambda$, 
\begin{equation}
e^{-6\lambda}= \frac{-\partial_yD}{y(1-y\partial_yD)} ,~~~~\Rightarrow~~~~ \frac{2V''}{\tilde{\Delta}} = \frac{-\partial_yD}{(1-y\partial_yD)} ,
\end{equation}
one arrives at
\begin{equation}\label{eqn:box3}
\partial_yD =\frac{2V''}{2V''\dot{V}-\tilde{\Delta}}=\frac{2V''}{\ddot{V}V''-(\dot{V}')^2},
\end{equation}
which matches the form in \cite{Reid-Edwards:2010vpm} after inserting the Laplace equation \eqref{eqn:laplace}.\\
\\Using $r=e^{V'}$ and $\sigma^2=r^2e^D$,
\begin{equation}\label{eqn:box4}
e^{\frac{1}{2}D}dr=  \sigma V''d\eta +\dot{V}'d\sigma.
\end{equation}
Hence,
\begin{equation}
dy^2+e^Ddr^2 =(d\sigma^2+d\eta^2)\bigg(-V''\ddot{V}+(\dot{V}')^2\bigg),
\end{equation}
meaning,
\begin{align}
-\frac{\partial_yD}{y}(dy^2+e^Ddr^2 )&= -\frac{1}{\dot{V}}\frac{2V''}{(\ddot{V}V''-(\dot{V}')^2)}\big(-V''\ddot{V}+(\dot{V}')^2\big)(d\sigma^2+d\eta^2)\nn\\[2mm]
&=f_4 (d\sigma^2+d\eta^2).
\end{align}
Thus, the only part of the LLM metric still left to transform is 
 \begin{equation}
\frac{4}{1-y\partial_yD}\Big(d\bar{\chi}-\frac{r}{2}\partial_rD d\bar{\beta}\Big)^2-\frac{\partial_yD}{y}e^D r^2d\bar{\beta}^2.
\end{equation}
Before doing so, we note the following relation from the transformation $\sigma(r,y)$ (as in \cite{Reid-Edwards:2010vpm})
\begin{equation}
\frac{d\sigma}{d\eta}=\frac{\partial \sigma}{\partial r} \frac{\partial r}{\partial \eta}+\frac{\partial \sigma}{\partial y} \frac{\partial y}{\partial \eta} = \partial_r\sigma \partial_\eta r+\partial_y \rho \partial_\eta y=0,
\end{equation}
giving 
\begin{equation}
\partial_r\sigma=-\frac{\partial_y \sigma \partial_\eta y}{ \partial_\eta r}.
\end{equation}
Now we use the forms of the transform to get (assuming $\sigma =re^{\frac{D}{2}}$)
\begin{equation}
\partial_r\sigma= \frac{1}{2}e^{\frac{D}{2}}(2+r\partial_r D),~~~~\partial_y\sigma= \frac{\sigma}{2}\partial_y D,~~~~\partial_\eta y = \dot{V}',~~~~\partial_\eta r = V'' r.
\end{equation}
Now, substituting these forms into the previous equation
\begin{equation}
\begin{aligned}
& e^{\frac{D}{2}}(2+r\partial_r D)=-\frac{\sigma\partial_y D \dot{V}'}{V'' r}\\
\Rightarrow~~~~~ & r\partial_r D=- \frac{\partial_y D \dot{V}'}{V'' }-2 = - \frac{2 \dot{V}'}{\ddot{V}V''-(\dot{V}')^2}-2 = -2\bigg[g(\eta,\sigma)+1\bigg].
\end{aligned}
\end{equation}
Note that it is this extra contribution to $g(\eta,\sigma)$ which introduces the additional $\beta$ term in the definition of the GM $\chi$.
Inserting this into the remaining part, gives
 \begin{equation}
\frac{4}{1-y\partial_yD}\Big(d\bar{\chi}-\frac{r}{2}\partial_rD d\bar{\beta}\Big)^2-\frac{\partial_yD}{y}e^D r^2d\bar{\beta}^2 
= \frac{4}{1-y\partial_yD}\bigg(d\chi-g(\eta,\sigma)d\beta\bigg)^2-\frac{\partial_yD}{y}e^D r^2d\beta^2,
\end{equation}
where 
\begin{equation}
\chi = \bar{\chi} + \bar{\beta},~~~~\beta=-\bar{\beta},~~~~g(\eta,\sigma) = \frac{ \dot{V}'}{\ddot{V}V''-(\dot{V}')^2}.
\end{equation}
Now we re-arrange this such that the roles of $\chi$ and $\beta$ are switched
\begin{align}
\frac{4}{1-y\partial_yD}\bigg(d\chi&-g(\eta,\sigma)d\beta\bigg)^2-\frac{\partial_yD}{y}e^D r^2d\beta^2= -\frac{4r^2e^D \partial_yD}{4y\,g(\eta,\sigma)^2 -e^D r^2\partial_yD (1-y\partial_yD)}d\chi^2\\[2mm]
+& \bigg(\frac{4y g(\eta,\sigma)^2 -r^2 e^D \partial_yD (1-y\partial_yD)}{y(1-y\partial_yD)}\bigg)\bigg[d\beta -\frac{4y\, g(\eta,\sigma)}{4y\,g(\eta,\sigma)^2 -e^D r^2 \partial_yD (1-y\partial_yD)}d\chi\bigg]^2\nn.
\end{align}
Noting 
\begin{equation}
4y g(\eta,\sigma)^2 -r^2 e^D \partial_yD (1-y\partial_yD) = -\frac{2\Lambda V''}{\ddot{V}V''-(\dot{V}')^2},~~~~~~~~~~y (1-y\partial_yD) = -\frac{\dot{V}\tilde{\Delta}}{\ddot{V}V''-(\dot{V}')^2},
\end{equation}
we transform each above form as follows
\begin{align}
\frac{4y g(\eta,\sigma)^2 -r^2 e^D \partial_yD (1-y\partial_yD)}{y(1-y\partial_yD)} = \frac{2\Lambda V''}{\dot{V} \tilde{\Delta}}&=f_5,\nn\\[2mm]
\frac{4y\, g(\eta,\sigma)}{4y\,g(\eta,\sigma)^2 -e^D r^2 \partial_yD (1-y\partial_yD)}= -\frac{2\dot{V}\dot{V}'}{\Lambda V''} &=-  f_6,\nn\\[2mm]
\frac{4r^2e^D \partial_yD}{4y\,g(\eta,\sigma)^2 -e^D r^2\partial_yD (1-y\partial_yD)} = -\frac{4\sigma^2}{\Lambda}&=-f_3.
\end{align}
Hence
\begin{align}
\frac{4}{1-y\partial_yD}\bigg(d\chi&-g(\eta,\sigma)d\beta\bigg)^2-\frac{\partial_yD}{y}e^D r^2d\beta^2\\
&= f_5\big(d\beta +f_6d\chi\big)^2+f_3d\chi^2.
\end{align}
For $A_3$, noting
\begin{equation}
2\kappa\, y^3 e^{-6\lambda} = -f_7,~~~~\kappa\, e^{-6\lambda}\frac{y^2}{\partial_yD}(2y\partial_yD-2g(\eta,\sigma)) = -f_7 - \kappa \frac{2\dot{V}\dot{V}'}{\tilde{\Delta}},~~~~\kappa\bigg(2e^{-6\lambda}\frac{y^2}{\partial_yD}+2y\bigg) = -f_7,\nn
\end{equation}
and 
\begin{equation}
- \partial_y(e^D)r\,dr +(2+r\partial_rD)\,dy =2d\eta,
\end{equation}
after taking the positive square-root for $\sigma = \pm r e^{\frac{D}{2}}$  
\begin{align}
G_4&= \kappa \bigg[-d(2y^3 e^{-6\lambda}) \wedge d\bar{\chi} -d(2y^3 e^{-6\lambda}) \wedge d\bar{\beta} \nn \\[2mm]
+&  \bigg(d\bigg[e^{-6\lambda}\frac{y^2}{\partial_yD}(2y\partial_yD+r\partial_rD)\bigg] - \partial_y(e^D)r\,dr + r\partial_rD\,dy\bigg) \wedge d\bar{\beta} \bigg]\wedge \text{vol}(\text{S}^2) \nn\\[2mm]
&=  \bigg[d(f_7) \wedge d\bar{\chi} +d(f_7) \wedge d\bar{\beta} \nn\\[2mm]
&+ \kappa \bigg(d\bigg[e^{-6\lambda}\frac{y^2}{\partial_yD}(2y\partial_yD-2g(\eta,\sigma)-2)\bigg] - \partial_y(e^D)r\,dr + r\partial_rD\,dy\bigg) \wedge d\bar{\beta} \bigg]\wedge \text{vol}(\text{S}^2) \nn\\[2mm]
&=  \bigg[d(f_7) \wedge d\bar{\chi} + d(f_7) \wedge d\bar{\beta}+  \bigg(d\bigg[ -f_7 -\kappa \frac{2\dot{V}\dot{V}'}{\tilde{\Delta}}\bigg] + 2\kappa\, d\eta -\kappa\, d\bigg[2e^{-6\lambda}\frac{y^2}{\partial_yD}+2y\bigg] \bigg) \wedge d\bar{\beta}  \bigg]\wedge \text{vol}(\text{S}^2) \nn\\[2mm]
&=  \bigg[d(f_7) \wedge (d\bar{\chi} +d\bar{\beta}) - d(f_8)  \wedge d\bar{\beta}  \bigg]\wedge \text{vol}(\text{S}^2).
\end{align}
Finally we arrive at the form of the GM solution written in eqn.\eqref{eqn:GM}. We note that $\chi$ in the GM paper is in fact $\chi = \bar{\chi} + \bar{\beta}$ written in terms of the LLM form (as stated in \cite{Bah:2019jts,Bah:2022yjf,Couzens:2022yjl}) (and $\bar{\beta}\rightarrow -\beta$).

\section{Values of $f_i$ at special points in space}\label{sec:valuesoffi}
In this appendix we quote the values of the functions $f_i$, which appear in our various solutions at special points in the space.\\
~\\
At the boundary $\sigma=\infty$ one finds to leading order that
\begin{align}
f_1&=\left(\frac{\pi^3 {\cal R}^2_1 \kappa^2 \sigma^2}{4 P^2}e^{-\frac{2\pi}{P}\sigma}\right)^{\frac{1}{3}},~~~f_2=\frac{2P}{\pi\sigma}\sin^2\left(\frac{\pi\eta}{P}\right),~~~f_3=4,~~~~f_4=\frac{2\pi}{P\sigma},~~~f_5=\frac{4P^2}{\pi^3 {\cal R}_1^2}e^{\frac{2\pi}{P}\sigma}\\[2mm]
f_6&=\sqrt{\frac{2}{P\sigma}}\pi {\cal R}_1 e^{-\frac{\pi}{P}\sigma}\cos\left(\frac{\pi\eta}{P}\right),~~~f_7=-2\kappa{\cal R}_1 \sqrt{\frac{2 P}{\sigma}}e^{-\frac{\pi}{P}\sigma}\sin^3\left(\frac{\pi \eta}{P}\right),~~~f_8=\kappa\left(-2 \eta+\frac{P}{\pi}\sin\left(\frac{2\pi\eta}{P}\right)\right)\nn.
\end{align}
At $\eta=0$ but $\sigma \neq 0$ the leading order behaviour 
\begin{align}
f_1&=\left(\frac{\kappa^2 \sigma^2 f^3}{-2\dot{f}}\right)^{\frac{1}{3}},~~~f_2=\frac{-2\eta^2\dot{f}}{\sigma^2 f},~~~f_3=\frac{-4 \dot{f}}{2f- \dot{f}},~~~f_4=-\frac{2 \dot{f}}{f\sigma^2},\\[2mm]
f_5&=\frac{2(2f -\dot{f})}{f^3},~~~f_6=\frac{2 f^2}{2f-\dot{f}},~~~f_7=\frac{4\kappa\eta \dot{f}}{\sigma^2},~~~f_8=2\kappa\left(-\eta+\frac{\eta}{f}\right),\nn
\end{align}
where $f>0$ is defined in \eqref{eq:fdef} and $-\dot{f}>0$. The behaviour at $\eta=P$ but $\sigma \neq 0$ is qualitatively the same. \\
~\\
To approach $\sigma=\eta=0$ we define $(\eta= r\cos\alpha,\sigma=r\sin\alpha)$ and expand in small $r$ leading to the asymptotic form
\begin{align}
f_1&=\left(\frac{\kappa^2 N_1^3 }{2 Q}\right)^{\frac{1}{3}},~~~f_2=\frac{2 r^2\cos^2\alpha Q}{N_1},~~~f_3= \frac{2 r^2 \sin^2\alpha Q}{N_1},~~~f_4=\frac{2 Q}{N_1},\\[2mm]
f_5&= \frac{4}{N_1^2},~~~f_6= N_1,~~~f_7=-4\kappa Q r^3 \cos^3\alpha,~~~f_8=0,\nn
\end{align}
where $Q$ can be extracted from \eqref{eqQdef} by decomposing $V''=r \cos\alpha Q$ - the behaviour at $\sigma=\eta-P=0$ is qualitatively the same.\\
~\\
The behaviour at $(\sigma=0,~\eta=k)$ for $0<k<P$ an integer where ${\cal R}'$ is discontinuous can be studied by defining $(\eta=k- r\cos\alpha,\sigma=r\sin\alpha)$ and expanding in small $r$, we find to leading order
\begin{align}
f_1&=(\kappa N_k)^{\frac{2}{3}},~~~f_2=1,~~~f_3=\frac{r^2\sin^2\alpha}{N_k}\frac{b_k}{r},~~~f_4=\frac{1}{N_k}\frac{b_k}{r},\\[2mm]
f_5&=\frac{4}{N_k}\frac{r}{b_k},~~~f_6=\frac{b_k}{2}(1+\cos\alpha)+N_{k+1}-N_k,~~~f_7=-2\kappa N_k,~~~f_8=-2\kappa k.\nn
\end{align} 
Finally at a generic regular point of the $\sigma=0$ boundary, for $\eta\in (k,k+1)$ for which ${\cal R}=N_k+(N_{k+1}-N_{k})(\eta-k)$  we find
\begin{align}
f_1&=\left(\frac{\kappa^{2}{\cal R}(2 {\cal R}P_k +({\cal R}')^2)}{2 P_k}\right)^{\frac{1}{3}},~~~f_2=\frac{2 {\cal R} P_k}{2 {\cal R} P_k+({\cal R}')^2},~~~f_3= \frac {2 \sigma^2 P_k}{\cal R},~~~f_4=\frac{ 2 P_k}{{\cal R}},\\[2mm]
f_5&=\frac{4}{2{\cal R}P_k+({\cal R}')^2},~~~f_6={\cal R}',~~~f_7=-\frac{4\kappa {\cal R}^2 P_k}{2{\cal R} P_k+({\cal R}')^2},~~~f_8=2\kappa\left(-\eta+\frac{{\cal R}{\cal R}'}{2{\cal R}P_k+({\cal R}')^2}\right),\nn
\end{align}
where $P_k=P_k(\eta)$ is defined in  \eqref{eqPdef} - note that if ${\cal R}'$ is actually continuous at either of $\eta=k$ or $\eta=k+1$ then these expressions also hold at that respective point.

\section{Some details on the $d=5$ KK monopole solution}\label{sec:KKdetials}
In this appendix we show how one can generate the flat space analogue of the KK monopole singularity we find in \eqref{eq:interestingmonopole} via coordinate transformations and dimensional reduction.\\
~~\\ 
One can  generate the relevant singular behaviour starting from $\mathbb{R}^6$ parameterised as
\beq
ds^2(\mathbb{R}^6)=  \frac{1}{\rho}\bigg(d\rho^2+4\rho^2 ds^2(\text{S}^5)\bigg)\label{eq:R6}.
\eeq
We first express S$^5$ as a foliation of S$^1\times$S$^3$ over an interval and then perform a $\mathbb{Z}_{p}$ orbifolding of the Hopf fiber coordinate within the 3-sphere, the result is
\beq
ds^2(\text{S}^5/\mathbb{Z}_{p})= d\mu^2+ \cos^2\mu d\phi^2+\frac{1}{4}\sin^2\mu(d\alpha^2+ \sin^2\alpha d\chi^2)+\frac{1}{p^2}\sin^2\mu(d\beta+ \frac{p}{2}\cos\alpha d\chi)^2,\label{eq:S5orbifold}
\eeq
which preserves half of the supersymmetry of the round 5-sphere in terms of two $d=5$ spinors $\zeta_{\pm}$, charged under respectively $\partial_{\phi}\pm \partial_{\chi}$ which are both singlets with respect to $\partial_{\beta}$. Specifically with respect to the obvious vielbein \eqref{eq:S5orbifold} suggests (taking plus signs) we find that
\beq
\zeta_{\pm}= e^{\frac{i}{2}(\phi\pm\chi)}e^{-\frac{i}{2}\mu \gamma_1}e^{\frac{1}{2}\alpha\gamma_{45}}\zeta^0_{\pm},
\eeq
both obey $\nabla_{a}\zeta_{\pm}+\frac{i}{2}\gamma_a \zeta_{\pm}=0$, 
where $\zeta^0_{\pm}$ are two constant and independent eigenvectors of $-\gamma_{15}+\gamma_{34}$ with zero eigenvalue.
 
Next one performs the following coordinate transformations in succession
\begin{align}
\beta &\to  \beta+ \frac{q}{2} \chi,\nn\\[2mm]
\chi&\to \chi+l\beta,~~~~\phi\to \phi-l \beta,
\end{align}
for $q,l$ constants, which leaves the spinor $\zeta_{+}$ invariant but makes $\zeta_{-}$ now charged under $\partial_{\beta}$ . The result of doing this is that $\text{S}^5/\mathbb{Z}_{p}\to \mathbb{ B}_5$, a U(1) fibration over a topological $\mathbb{CP}^2$ with orbifold singularities, given specifically by
\begin{align}
ds^2(\mathbb{ B}_5)&= ds^2(\mathbb{B}^{pql}_4)+\frac{\Xi_{plq}}{p^2}(d\beta+{\cal B}_1)^2,~~~~ds^2(\mathbb{B}^{pql}_4)=d\mu^2\nn\\[2mm]
+&\frac{1}{4}\sin^2\mu\left(d\alpha^2+\frac{\sin^2\alpha}{\Delta_{pql}}d\chi^2\right)+\frac{\Xi_{pql}}{p^2}+ \frac{\Delta_{pql}}{\Xi_{pql}}\sin^2\mu\cos^2\mu\left(d\phi+ \frac{\Delta_{pql}-1-\frac{l}{2}(q+p\cos\alpha)}{\Delta_{pql}}d\chi\right)^2,\nn\\[2mm]
{\cal B}_1&=\frac{-p^2l \cos^2\mu d\phi+\frac{1}{\xi}(\Delta-1-\frac{l}{2}(q+p \cos\alpha))\sin^2\mu d\chi}{\Xi_{pql}},\nn\\[2mm]
\Xi_{pql}&=\sin^2\mu\Delta_{pql}+p^2l^2\cos^2\mu,~~~~\Delta_{pql}=\frac{1}{4}p^2l^2\sin^2\alpha+\left(1+\frac{l}{2}(q+p\cos\alpha)\right)^2.
\end{align}
If we replace S$^5$ with $\mathbb{ B}_5$ in \eqref{eq:R6} we end up with a rather complicated orbifolding of $\mathbb{R}^6$, that we propose is the near horizon limit of some more complicated $d=6$ monopole geometry that is likewise Ricci flat, \textit{i.e} a CY$_3$ analogy of the geometries constructed in \cite{Gauntlett:1997pk}, whose form we do not know. Taking the direct product of $\mathbb{R}^{1,4}$ and this $\mathbb{R}^6$ orbifold yields a purely geometric solution to $d=11$ supergravity which we are free to reduce to type IIA on the isometry $\partial_{\beta}$, which preserves only the $\zeta_+$ spinor, halving supersymmetry again. The resulting solution in IIA takes the form
\beq
ds^2=\sqrt{\Xi}_{pql}\bigg[\frac{1}{\sqrt{\frac{p}{\rho}}}ds^2(\mathbb{R}^{1,5})+\sqrt{\frac{p}{\rho}}\left(d\rho^2+4\rho^2 ds^2(\mathbb{B}_4^{pql})\right)\bigg],~~~e^{-\Phi}=\left(\frac{l^6}{2^6\Xi_{pql}\rho^3}\right)^{\frac{1}{4}},
\eeq
with non trivial RR 2-form $F_2=d{\cal B}_1$. This yields an object in IIA with the same behaviour as \eqref{eq:interestingmonopole}, but now extended in flat space - specifically one should identify $(p=b_k~,q=N_{k+1}-N_{k-1}~l=\xi)$ upon which $\Delta_{pql}\to \Delta_k$ and $\Xi_{pql}\to \tilde{\Xi}_k$.

 \section{$\mathcal{N}=1$ G-Structures}
In this section we present the G-structure that the ${\cal N}=1$ preserving deformation discussed in section \ref{sec:neq1sol} preserves.\\
~\\
The G-structure forms for the  ${\cal N}=1$ case can be extracted in a similar fashion to those of ${\cal N}=2$ as performed in section \ref{sec:g-structures}. After performing the appropriate shifts in the U(1) isometry directions, the vielbein frame of \eqref{eqn:GstructureForms} gets modified to
 \begin{align}
 K&=\frac{\kappa e^{-2\rho}}{f_1}d(\cos\theta e^{2\rho}\dot{V}),\nn\\[2mm]
 E_1
&=- \sqrt{ \frac{f_1f_3 }{\Xi \,\Sigma_1 }} \Bigg[ \frac{1}{\sigma}d\sigma+ d\rho+ i\,(\Sigma_1\, d\chi +\xi \,\Sigma_2\sin^2\theta d\phi)\nn \\[2mm]
&+\xi\bigg(d(V') +\frac{e^{-2\rho}}{2\dot{V}}\Big(
\Sigma_2 \, d(e^{2\rho}\dot{V}\sin^2\theta\big)+\Big(\Sigma_2- \xi \frac{4f_2^2}{f_3f_5}\Big)\sin^2\theta d(e^{2\rho}\dot{V}) +\xi \frac{4f_2}{f_3f_5} \sin^2\theta \,\dot{V}d(e^{2\rho} )\Big) \bigg)\Bigg],\nn\\[2mm]
 E_2&=\frac{ e^{i \phi }}{\sqrt{\Sigma_1}} \Bigg[\frac{\kappa}{f_1 }\Big(e^{-2\rho}d( e^{2\rho}\dot{V}\sin\theta)  -\xi \,\dot{V} d(V') \sin\theta \Big)+i\, \sqrt{f_1f_2}\sin\theta\,d\phi \Bigg],\nn\\[2mm]
 E_3&=-e^{i\chi }\,\Xi^{\frac{1}{2}} \sqrt{f_1f_5}  
 \Bigg[ 
 \frac{1}{\Xi}\left(-\frac{f_3 \dot{V}'}{4 \sigma}d\sigma -V''d\eta+f_6d\rho + \frac{\xi}{f_5}\Big(\frac{\kappa^2e^{-4\rho}}{2f_1^3}d(e^{4\rho}\dot{V}^2(\cos^2\theta-3))+4 d\rho\Big)\right)\nn\\[2mm] 
 &+i\big(d\beta + C_1\big)\Bigg],
\end{align}
where we define
\begin{align}
\Sigma_1 &= 
1+\xi^2\frac{f_2}{f_3}\Big(f_6^2 +\frac{f_3}{f_5}\Big)\sin^2\theta,~~~~~~~~~~~\Sigma_2= \frac{f_2}{f_3}\bigg(f_6(1+\xi f_6)+\xi \frac{f_3}{f_5}\bigg),~~~~~ ~~~~  e^{\frac{4}{3}\Phi}=  f_1 f_5\Xi ,        \nn  \\[2mm]
\Xi&=(1+ \xi f_6)^2+\xi^2\frac{f_3}{f_5}+\xi^2\frac{f_2}{f_5}\sin^2\theta,~~~~
        C_1= \frac{1}{\Xi}\bigg(\Big(f_6+\xi\Big(f_6^2+\frac{f_3}{f_5}\Big)\Big)d\chi-\xi \frac{f_2}{f_5}\sin^2\theta d\phi\bigg),
        \end{align}
and $(y_1\equiv \cos\phi\sin\theta, y_2\equiv \sin\phi\sin\theta,y_3\equiv \cos\theta)$. The SU(2) structure forms can be extracted from this expression following the formulae in section \ref{sec:g-structures}, we find they are given by 
\begin{align}
v&=  \kappa e^{-2\rho}f_5^{\frac{1}{4}}f_1^{-\frac{3}{4}} \Xi^{\frac{1}{4}}d\left(e^{2\rho} \dot{V}y_3\right),\nn\\[2mm]
u&= (f_1f_5)^{\frac{3}{4}}\Xi^{-\frac{1}{4}}\left(\frac{f_3 \dot{V}'}{4 \sigma}d\sigma +V''d\eta-f_6d\rho - \frac{\xi}{f_5}\Big(\frac{\kappa^2e^{-4\rho}}{2f_1^3}d(e^{4\rho}\dot{V}^2(y_3^2-3))+4 d\rho\Big)\right),\nn\\[2mm]
\omega&=  -2(\kappa)^2  f_1^{-\frac{3}{2}}e^{-3\rho}d\bigg(e^{2\rho}\dot{V} e^{-\xi V'}(y_1+i y_2)d(e^{i\chi}e^{\rho}e^{\xi V'}\sigma)\bigg),\nn\\[2mm]
j&=\frac{1}{\sqrt{\Xi}}\Bigg[ \frac{f_1^{\frac{3}{2}}f_5^{\frac{1}{2}} f_3}{\sigma} e^{-\rho}e^{-\xi V'}d\left(e^{\rho}\sigma e^{\xi V'}\right)\wedge d\chi+ \kappa f_2 f_3^{\frac{1}{2}} X_1\wedge (d\chi+d\phi)\label{eq:neq1iiagstructureforms}\\[2mm]
&+\kappa\Bigg(\frac{\sigma f_2 e^{-4\rho}}{f_3^{\frac{1}{2}} \dot{V}^2}(1+\xi f_6)d(e^{4\rho}\sin^2\theta \dot{V}^2)-\frac{\xi}{2} f_2f_3^{\frac{1}{2}}\sin^2\theta\left(\dot{V}'\left(d\sigma-\frac{2 \dot{V}}{\sigma V''}d\rho\right)+\frac{d\eta}{\sigma}\left(2 \dot{V}- \ddot{V}\right)\right)\Bigg)\wedge d\phi\Bigg]\nn,\\[2mm]
X_1&=\xi \frac{\dot{V}' e^{-4\rho}}{2 \dot{V} V''\sigma}d(e^{4\rho}\sin^2\theta\dot{V}^2)+\xi^2\frac{(-V'' \ddot{V}+(\dot{V}')^2 )e^{-6\rho}}{2\sigma \dot{V} V''}d(e^{6\rho}\sin^2\theta\dot{V}^2)+\xi^2 \dot{V}\sin^2\theta\Big(V'' d\sigma-\frac{\dot{V}'}{\sigma}d\eta\Big).\nn
\end{align}
 with $z=u+iv$. The Pure Spinors are given as before by \eqref{eqn:Psipm}.
 

  \section{Holographic Central Charge}\label{sec:HCC}         
For the general Holographic Central Charge, $c_{hol}$, 
\begin{equation}
\begin{gathered}
ds^2=\alpha(\rho,\vv{\theta})\Big(dx_{1,d}^2+\beta(\rho)d\rho^2\Big)+g_{ij}(\rho,\vv{\theta})d\theta^id\theta^j,\\
c_{hol}=\frac{d^d}{G_N}\beta^{d/2}\frac{H^{\frac{2d+1}{2}}}{(H')^d},~~~~H=V_{int}^2,~~~~~V_{int}=\int d\vv{\theta}\sqrt{\text{det}[g_{ij}]e^{-4\Phi}\alpha^d},
\end{gathered}
\end{equation}
with $G_N=8\pi^6 \alpha'^4g_s^2=8\pi^6$. In the present cases, $d=3$.\\
\\All of the type IIA backgrounds encountered throughout the paper have a metric which can be expressed in the same manner, with all dependence on the parameters dropping out of the following calculation. In the simplest case, we have 
\begin{align}
    ds_{10,st}^2&=4\rho^2e^{\frac{2}{3}\Phi}f_1(\sigma,\eta)\bigg(dx_{1,3}^2+\frac{1}{\rho^4}d\rho^2\bigg)+ e^{\frac{2}{3}\Phi}f_1(\sigma,\eta)\bigg[f_2(\sigma,\eta)ds^2(\text{S}^2)+f_4(\sigma,\eta)(d\sigma^2+d\eta^2)\bigg]\nn\\[2mm]
    &+e^{-\frac{2}{3}\Phi}f_1(\sigma,\eta)^2f_5(\sigma,\eta)f_3(\sigma,\eta) d\chi^2.
\end{align}
It is now easy to read off the following 

\begin{equation}
\sqrt{\text{det}[g_{ij}]e^{-4\Phi}\alpha^3} = 8\rho^3 f_1^{\frac{9}{2}}f_3^{\frac{1}{2}}f_5^{\frac{1}{2}} f_2f_4  \Vol(\text{S}^1)\Vol(\text{S}^2)  ,
\end{equation}
which is independent of $\Phi$, meaning that all the backgrounds in this paper have the above form for the Holographic Central Charge!
\\\\Inserting the forms of the Warp Factors given in equations \eqref{eq:thefs}, gives
\begin{equation}
V_{int}=2^5\kappa^3 \Vol(\text{S}^1)\Vol(\text{S}^2) \,\rho^3  \int \sigma \dot{V} V''  d\sigma d\eta    ,
\end{equation}
leading to (with $ \Vol(\text{S}^1)=2\pi,~\Vol(\text{S}^2) =4\pi$)

\begin{equation}
\begin{gathered}
c_{hol}=\frac{3^3}{8\pi^6}\bigg(\frac{1}{\rho^4}\bigg)^{3/2}\frac{V_{int}^7}{[(V_{int}^2)']^3}  = \frac{4\kappa^3}{\pi^4} \int \sigma \dot{V} V''  d\sigma d\eta,  
\end{gathered}
\end{equation}
where $(V_{int}^2)'$ is the derivative of $V_{int}^2$ with respect to $\rho$, 
\beq
(V_{int}^2)'=\frac{6}{\rho}V_{int}^2.
\eeq
One can calculate this via two slightly different approaches. First, we can integrate directly in $\sigma$
\beq
\begin{aligned}
\int \sigma \dot{V} V''  d\sigma d\eta   = -\int_0^P d\eta \int_0^\infty  \dot{V}\partial_\sigma(\dot{V})d\sigma =-\frac{1}{2}\int_0^P  \Big[\dot{V}^2\Big]_0^\infty d\eta.
\end{aligned}
\eeq
Now, using the following 
\beq
\dot{V}(\sigma,\eta) = \sigma \sum_{k=1}^\infty \,a_k \sin\bigg(\frac{k\pi}{P}\eta\bigg) K_1\bigg(\frac{k\pi}{P}\sigma\bigg),~~~~\lim_{x\rightarrow \infty} x K_1(x) \sim \sqrt{\frac{\pi}{2}}\sqrt{x}e^{-x} = 0,~~~~\lim_{x\rightarrow 0} x K_1(x)=1,\nn
\eeq
leads to
\begin{align}
\dot{V}^2 &= \sum_{k=1}^\infty \sum_{j=1}^\infty a_ka_j \bigg[\sigma K_1\bigg(\frac{k\pi}{P}\sigma\bigg)\bigg]\bigg[\sigma K_1\bigg(\frac{j\pi}{P}\sigma\bigg)\bigg]\sin\bigg(\frac{k\pi}{P}\eta\bigg)\sin\bigg(\frac{j\pi}{P}\eta\bigg)\bigg|_{\sigma=0}^{\sigma=\infty}\nn\\[2mm]
 &= \sum_{k=1}^\infty \sum_{j=1}^\infty a_ka_j \frac{P^2}{kj\pi^2}\bigg[\frac{k\pi \sigma}{P} K_1\bigg(\frac{k\pi}{P}\sigma\bigg)\bigg]\bigg[\frac{j\pi \sigma}{P} K_1\bigg(\frac{j\pi}{P}\sigma\bigg)\bigg]\sin\bigg(\frac{k\pi}{P}\eta\bigg)\sin\bigg(\frac{j\pi}{P}\eta\bigg)\bigg|_{\sigma=0}^{\sigma=\infty}\nn\\[2mm]
  &= - \sum_{k=1}^\infty \sum_{j=1}^\infty a_ka_j \frac{P^2}{kj\pi^2}\sin\bigg(\frac{k\pi}{P}\eta\bigg)\sin\bigg(\frac{j\pi}{P}\eta\bigg),
\end{align}
meaning
\begin{align}
c_{hol} = \frac{4\kappa^3}{\pi^4}  \int \sigma \dot{V} V''  d\sigma d\eta  &= \frac{2\kappa^3}{\pi^4}  \int_0^P   \sum_{k=1}^\infty \sum_{j=1}^\infty a_ka_j \frac{P^2}{kj\pi^2}\sin\bigg(\frac{k\pi}{P}\eta\bigg)\sin\bigg(\frac{j\pi}{P}\eta\bigg)   d\eta\nn\\[2mm]
&= \frac{\kappa^3 P^3}{\pi^6}    \sum_{k=1}^\infty  \frac{a_k^2 }{k^2}\nn \\[2mm]
&=\frac{ \kappa^3}{\pi^4}\sum_{k=1}^\infty P \, \mathcal{R}_k^2,
\end{align}
after using 
\begin{equation} 
 a_k =\frac{k\pi}{P}\mathcal{R}_k,~~~~   \int_0^P \sin\bigg(\frac{k\pi}{P}\eta\bigg)\sin\bigg(\frac{j\pi}{P}\eta\bigg) d\eta =\frac{P}{2}\delta_{kj}.
    \end{equation}
Alternatively, one can insert the following definitions into the initial form, $\int \sigma \dot{V} V''  d\sigma d\eta$,
\begin{equation}
\begin{aligned}
V(\sigma,\eta) &=-\sum_{k=1}^\infty \bigg(\frac{P}{k\pi}\bigg)\,a_k \sin\bigg(\frac{k\pi}{P}\eta\bigg) K_0\bigg(\frac{k\pi}{P}\sigma\bigg),\\
\dot{V}(\sigma,\eta) &= \sigma \sum_{k=1}^\infty \,a_k \sin\bigg(\frac{k\pi}{P}\eta\bigg) K_1\bigg(\frac{k\pi}{P}\sigma\bigg),\\
V''(\sigma,\eta) &=\sum_{k=1}^\infty \bigg(\frac{k\pi}{P}\bigg)\,a_k \sin\bigg(\frac{k\pi}{P}\eta\bigg) K_0\bigg(\frac{k\pi}{P}\sigma\bigg),\\
\sigma \dot{V}(\sigma,\eta)V''(\sigma,\eta) &= \sigma^2 \sum_{k=1}^\infty\sum_{l=1}^\infty \,a_k a_l \bigg(\frac{k\pi}{P}\bigg) \sin\bigg(\frac{k\pi}{P}\eta\bigg) \sin\bigg(\frac{l\pi}{P}\eta\bigg) K_0\bigg(\frac{k\pi}{P}\sigma\bigg) K_1\bigg(\frac{l\pi}{P}\sigma\bigg),
\end{aligned}
\end{equation}
using the following results 
  \begin{equation}\label{eqn:standard}
    \int_0^P \sin\bigg(\frac{k\pi}{P}\eta\bigg)\sin\bigg(\frac{l\pi}{P}\eta\bigg) d\eta =\frac{P}{2}\delta_{kl},~~~  \int_{0}^\infty \sigma^2 K_0\bigg(\frac{k\pi}{P}\sigma\bigg)K_1\bigg(\frac{k\pi}{P}\sigma\bigg) d\sigma = \frac{P^3}{2\pi^3k^3} ,
    \end{equation}
which leads to the same result as the first approach (with $ a_k =\frac{k\pi}{P}\mathcal{R}_k$)
\begin{equation}
\begin{aligned}
c_{hol} &= \frac{4\kappa^3}{\pi^4} \int \sigma \dot{V} V''  d\sigma d\eta =\frac{ \kappa^3 P^3 }{\pi^6}\sum_{k=1}^\infty \frac{a_k^2}{k^2}\\
&=\frac{ \kappa^3}{\pi^4}\sum_{k=1}^\infty P \, \mathcal{R}_k^2.
\end{aligned}
\end{equation}

\end{document}